\documentclass{article}% Math and Physical Sciences Reference Style

\usepackage{hyperref}
\usepackage[latin1]{inputenc}
\usepackage{algorithm}

\usepackage{algpseudocode}
\usepackage{caption}
\usepackage[binary-units]{siunitx}
\usepackage{booktabs}
\usepackage{multirow}
\usepackage[export]{adjustbox}
\usepackage{subcaption}

\algblock[Name]{Grover}{EndGrover}

\usepackage{amssymb,latexsym,amsmath}
\usepackage{amsthm}
\usepackage{physics}
\usepackage{graphicx}
\usepackage{graphics}
\usepackage[T1]{fontenc}
\usepackage{mdframed}
\usepackage{textcomp}

\newtheorem{remark}{Remark}%

\newtheorem{definition}{Definition}%

\title{On recovering block cipher secret keys in the cold boot attack setting}

\author{Gustavo Banegas\\
Inria and Laboratoire d'Informatique de l'Ecole polytechnique\\ Institut Polytechnique de Paris Palaiseau, France.\\ \\ Ricardo Villanueva-Polanco \\ Department of Computer Science and Engineering \\Universidad del Norte, Barranquilla, Colombia.
}

\date{}

\begin{document}

\maketitle

%\author{\fnm{Gustavo} \sur{Banegas}}\email{gustavo@cryptme.in}

%\author*[2]{\fnm{Ricardo} \sur{Villanueva-Polanco}}\email{rpolanco@uninorte.edu.co}


%\affil[1]{\orgdiv{
%Inria and Laboratoire d'Informatique de l'Ecole polytechnique}, %\orgname{Institut Polytechnique de Paris}, \orgaddress{ \city{Palaiseau}, \country{France}}}

%\affil*[2]{\orgdiv{Department of Computer Science and Engineering}, \orgname{Universidad del Norte}, \orgaddress{\street{KM 5 Via Puerto Colombia}, \city{Barranquilla}, \postcode{081007}, \state{Atl\'antico}, \country{Colombia}}}

%\affil[3]{\orgdiv{Department}, \orgname{Organization}, \orgaddress{\street{Street}, \city{City}, \postcode{610101}, \state{State}, \country{Country}}}


\abstract{This paper presents a general strategy to recover a block cipher secret key in the cold boot attack setting. More precisely, we propose a key-recovery method that combines key enumeration algorithms and Grover's quantum algorithm to recover a block cipher secret key after an attacker has procured a noisy version of it via a cold boot attack.  We also show how to implement the quantum component of our algorithm for several block ciphers such as AES, PRESENT and GIFT, and LowMC. Additionally, since evaluating the third-round post-quantum candidates of the National Institute of Standards and Technology (NIST) post-quantum standardization process against different attack vectors is of great importance for their overall assessment, we show the feasibility of performing our hybrid attack on Picnic, a post-quantum signature algorithm being an alternate candidate in the NIST post-quantum standardization competition. According to our results, our method may recover the Picnic private key for all Picnic parameter sets, tolerating up to $40\%$ of noise for some of the parameter sets.  Furthermore, we provide a detailed analysis of our method by giving the cost of its resources, its running time, and its success rate for various enumerations.}




\section{Introduction}

Post-quantum cryptography has gained much attention in the past few years. One of the main reasons is the National Institute of Standards and Technology (NIST) call for proposals for post-quantum schemes (Signature schemes and Key encapsulation mechanisms). Currently, the call is in the third round, and there are few candidates for signature schemes: Picnic, Falcon, Rainbow, Crystals-Dilithium, GeMSS, and SPHINCS+.

The security of the schemes relies on different mathematical properties, so one can break a scheme if one finds a way to exploit some weaknesses in these mathematical properties, and hence may, in an easy way, recover information that is sensitive. Moreover, there are attacks where the main target is the implementation of the scheme, and such attacks are called side-channel attacks. One of those attacks is called a cold boot attack. Briefly, the idea of the attack is to fetch sensitive data from the memory of an electronic device.

This paper presents a general procedure by which a cold boot attacker may recover a block cipher secret key after procuring a noisy version of the key via a cold boot attack. More specifically, we describe a method that exploits key enumeration algorithms and a well-known quantum algorithm, namely, Grover's Algorithm. Also, we show how to implement the quantum component of our algorithm for several block ciphers such as AES, PRESENT and GIFT, and LowMC. Furthermore, we give a use case where Picnic (a third-round signature scheme from NIST) is evaluated in the cold boot attack setting, focusing on its current reference implementation. According to our knowledge, this is the first paper evaluating this signature scheme in the cold boot attack setting. In the study case, we further detail our key-recovery method for Picnic private keys in the cold boot attack setting, providing a detailed analysis of its costs of resources, its running time, and success rates for all Picnic parameter sets.

This paper is structured as follows. In Section~\ref{sec:background}, we present background material about cold-boot attacks, the model we assume for studying cold-boot attacks on cryptographic schemes, as well as a literature review on previous works on cold-boot attacks on cryptographic algorithms and background material on quantum computing. Section~\ref{GAKR} gives a high-level idea of the key-recovery problem in the cold blood attack setting.  In Section~\ref{sec:recovering}, we present our hybrid key-recovery method. In particular, Section~\ref{sec:hybridattack} describes our key-recovery strategy combined with Grover's quantum algorithm (a.k.a hybrid attack), its running time, and costs in terms of resources for several block ciphers. In Section~\ref{sec:Picnic}, we concentrate on Picnic, particularly on its key-generation algorithm and implementation, providing a detailed description of how to apply our algorithm to LowMC in the context of Picnic.  Lastly,  Section~\ref{sec:conclusions} encloses our final comments on the paper, highlighting some future research works.

\label{sec:introduction}

\section{Background}
\label{sec:background}
In this section, we will present background material about cold boot attacks, the model we assume for studying cold-boot attacks on cryptographic schemes, a literature review of previous works about cold boot attacks on cryptographic algorithms, background material on quantum computing, and lastly, a general strategy to tackle the key-recovery problem in the cold boot attack setting.

\subsection{Cold boot attacks}
\label{subsec:cold_boot}

A cold boot attack is a kind of data remanence attack by which an adversary could fetch sensitive data from an electronic device's main memory after the device has supposedly deleted the memory data. This attack vector exploits the data remanence property of Dynamic RAM (DRAM). Through it,  an adversary might recover readable memory content after the device's power is off for a while.  This attack vector, introduced in~\cite{USENIX:HSHCPCFAF08}, has been explored extensively against multiple cryptographic schemes, as we will discuss in Section~\ref{CBAPW}.  In this setting, an adversary, who has physical access to a device, might retrieve chunks of memory content from the device via carrying out a cold-rebooting on it~\cite{USENIX:HSHCPCFAF08,7299928,9260613}. In general terms, the adversary forces the operating system to shut down, which causes it to go past all tasks that typically execute during a normal shutdown, such as the file system synchronization. Therefore such an adversary may employ an external disk to start and run a lightweight operating system to copy memory contents of pre-boot DRAM to a file. Alternatively, such an attacker may remove the physical memory modules from the device (if possible) and place them in an adversary-controlled device. The attacker then may run a lightweight operating system to copy and paste chunks of memory content from these physical memory modules to an external drive.  Because of some physical effects on the main memory,  the memory bits experience a deterioration process once the device's power is off, by which some bits get changed. Particularly some $0$ bits of the original content change to $1$ bits and vice-versa. Therefore the extracted data from the target device's main memory will be recognizably different from the original memory data.

Previous works~\cite{USENIX:HSHCPCFAF08,7299928,9260613} point out that an attacker can decelerate the bit degrading process by means of spraying a chemical product, like liquid nitrogen, onto the memory modules (that is, spraying cold compressed liquid onto the modules may maintain the original bit states for a prolonged period). Nonetheless, the attacker has yet to extract the memory content before restoring any important information from the target device's main memory. To extract chunks of memory, the attacker has to handle several possible issues. On rebooting, the initial boot process may overwrite chunks of memory with its running code and data, even though the overwritten chunks are normally small. Moreover, the initial boot process might execute a destructive memory check, yet this memory check may be bypassed. In particular, the attacker may use memory-imaging tools to produce correct dumps of memory contents to any external device, as was reported in~\cite{USENIX:HSHCPCFAF08,7299928,9260613}. These tools consume trivial amounts of RAM and usually are placed in memory in such a way that do not affect the data of interest. In case that such an attacker cannot force boot memory-imaging tools, the attacker removes the memory modules and place them in a compatible device and copy and paste the content to an external disk, like mentioned by the authors of~\cite{USENIX:HSHCPCFAF08}.

Once the attacker extracts some memory content, the attacker has to profile the content to estimate the probabilities of bit-flipping. That is the probability for a $1$ to $0$ bit flipping and a $0$ to $1$ bit flipping. Furthermore, according to the results of the experiment reported in~\cite{USENIX:HSHCPCFAF08}, almost all memory bits tend to decay to predictable ``ground'' states, with only a  portion flipping in the opposite direction. Additionally, the authors of ~\cite{USENIX:HSHCPCFAF08} mention that the probability of a bit-flipping in the opposite direction stays constant and is very small (circa $0.01$) as time elapses, while the probability for a bit to decay to the ground state increases over time. These results suggest that the attacker could model the decay in a portion of the memory as a binary asymmetric channel, i.e., we can assume that the probability for a $1$ to $0$ bit flipping is a fixed number and that the probability for a $0$ to $1$ bit flipping is another fixed number in a given time. Note that by reading and counting the number of $0$ bits and $1$ bits, the attacker can discover the ground state of a specific memory region. Additionally, the attacker can estimate the bit-flipping probabilities by comparing the bit count of original content in a memory region with its corresponding noisy version.

Finding encryption keys after procuring memory content is another challenge that the attacker has to address. Such a problem has been extensively discussed in~\cite{USENIX:HSHCPCFAF08} for Advanced Encryption Standard (AES) and RSA keys in-memory images. Even though the algorithms presented in~\cite{USENIX:HSHCPCFAF08} are scheme-specific, their algorithmic rationale may be easily adapted to devise key-finding algorithms for other schemes. These algorithms search for specific secret-key-identifying characteristics in the secret key in-memory formats as identifying labels for sequences of bytes. More precisely, these algorithms search for byte sequences with low Hamming distance to these identifying labels and verify that the remaining bytes in a possible sequence satisfy some conditions. Once the previous issues are coped with, the attacker will obtain a version with errors of the original secret key obtained from the memory image. Hence the attacker's ultimate goal is to reconstruct the original private key from its noisy version with the help of public cryptographic data associated with the target key.

The study of cold boot attacks on cryptographic algorithms has focused on developing key-recovery algorithms to efficiently and effectively reconstruct a secret key from its noisy version with the help of associated public cryptographic data for a target cryptosystem and evaluate the robustness and tolerance of these key-recovery algorithms to noise.

\subsection{Cold boot attack model}\label{CBAM}

Based on our previous discussion on cold boot attacks, we assume an attacker knows about the data structures storing the private key in memory and has access to the corresponding public parameters without any noise.  Also, we suppose such an attacker procures a noisy version of the target private key via applying several key finding algorithms. We note that finding the memory region that stores the private key is require to carry out this attack in practice and may be taken care of via applying several key finding algorithms~\cite{USENIX:HSHCPCFAF08,7299928,9260613}. Therefore, the adversary's main objective is to reconstruct the original private key.

We denote $\alpha = P(0 \rightarrow 1)$ as the probability of a $0$ to $1$ bit-flipping (a $0$ bit in the bit representation of the private key changes to a $1$ bit). Moreover, we denote $ \beta = P(1 \rightarrow 0)$ as the probability of a $1$ to $0$ bit-flipping (viz. a $1$ bit in the bit representation of the secret key changes to a $0$ bit). Furthermore, based on experimental results obtained in~\cite{USENIX:HSHCPCFAF08,7299928,9260613}, we assume one of these values  is very small (approximately $0.001$) and not liable to variation over time, while the other value does increase
over time.
As stated by preceding works on cold boot attacks~\cite{USENIX:HSHCPCFAF08,7299928,9260613}, such an attacker may estimate both $\alpha$ and $\beta$ by comparing original content with its corresponding noisy version (using the public key), and both remain fixed across the memory region that stores the private key.

% Reescribir
\subsection{Literature review }\label{CBAPW}

In this section, we present a review of previous works about cold boot attacks on cryptographic schemes.  In particular, we introduce this literature review by describing cold boot attacks on RSA, then cold boot attacks on discrete-logarithm-based schemes, then cold boot attacks on symmetric-key schemes, and finally cold boot attacks on post-quantum schemes.

\subsubsection{RSA setting}

The research paper by Heninger and Shacham~\cite{C:HenSha09} is the first work dealing with this class of attacks on RSA keys. They introduce a key-recovery algorithm, which relies on Hansel lifting and exploit the redundancy found in the popular RSA secret key in-memory format. The research papers by Henecka et al.~\cite{C:HenMayMeu10} and Paterson et al.~\cite{AC:PatPolSib12} further the initial work, and both research papers exploit the mathematical structure on which RSA relies. Furthermore, the research paper by Paterson et al.~\cite{AC:PatPolSib12} further concentrates on the error channel's asymmetric nature, which is intrinsically connected to the cold boot setting, analyzing the key-recovery problem from an information-theoretic perspective.

%Heninger and Shacham~\cite{C:HenSha09} were the first to explore the case for RSA keys. They introduced an efficient algorithm based on Hensel lifting that exploits redundancy in the typical RSA private key format. This work was followed up by Henecka, May and Meurer~\cite{C:HenMayMeu10} and Paterson, Polychroniadou and Sibborn~\cite{AC:PatPolSib12}, with both research papers also paying particular attention to the mathematically highly structured RSA setting. The research paper by Paterson et. al. in particular identified the asymmetric nature of the error channel intrinsic to the cold boot setting and revisit the problem of key recovery for cold boot attacks in an information theoretic manner.

\subsubsection{Discrete logarithm setting}

The first research work that looks into this attack in the discrete logarithm setting is by Lee et al.~\cite{CISC:LKBC11}. This work pays particular attention to recovering the secret key $x$ given the public key $g^x$, where $g$ is a field element and $x$ is a positive integer. Their model assumes the attacker has access to the public key $g^x$ and the noisy version of the private key $x$, as well as knowledge of an upper bound on the number of errors found in the noisy version of the secret key. Since their algorithm assumes knowing such an upper bound (hardly achievable) and exploits small redundancy in the secret-key format, it does not perform well in recovering keys if these keys are susceptible to considerable levels of noise.

A follow-up work by Poettering and Sibborn~\cite{RSA:PoeSib15} also explores this attack in the discrete logarithm setting, more concretely in the elliptic curve cryptography (ECC) setting. Their work is practical since it centers on two implementations for elliptic curve cryptography. In particular, this work exploits redundancy present in two secret key in-memory formats from two popular ECC implementations from Transport Layer Security (TLS) libraries.  They develop a dedicated key-recovery algorithm in the bit-flipping model for each studied memory representation, showing better results than the preceding work.

%were the first to explore these attacks in the discrete logarithm setting. They assumed an attack model in which the adversary has knowledge of the public key $g^x$, a noisy version of the private key $x$ and an upper bound for the number of errors in the private key. Since this latter assumption may not be realistic and the attacker did not have access to further redundancy, their proposed algorithm would likely fail to recover keys that were affected by particularly high noise levels in the true cold boot scenario, i.e., only assuming a bit-flipping model. This work was followed upon by Poettering and Sibborn~\cite{RSA:PoeSib15}, who reviewed two practical elliptic curve cryptography implementations to find exploitable in-memory representations. In particular, they focused on two scenarios taken from two ECC implementations from TLS libraries: the windowed non-adjacent form (wNAF) representation used in OpenSSL, and the comb-based approach used in PolarSSL. By exploiting redundancy found in the respective in-memory private key representations, cold boot key-recovery algorithms were developed and tested in the true cold boot scenario.

\subsubsection{Symmetric key setting}

Regarding the feasibility of cold boot attacks against symmetric-key primitives, several papers have already explored this class of attacks against some prominent block ciphers. At first, the paper by Albrecht and Cid~\cite{ACNS:AlbCid11} concentrates on the recovery of symmetric encryption keys by employing polynomial system solvers. Particularly, they use integer programming techniques and apply them to the key-recovery of Serpent block cipher's secret keys, and also introduce a dedicated key-recovery algorithm to Twofish secret keys. Furthermore, the paper by Kamal and Youssef~\cite{Kamal:2010} introduces key-recovery algorithms based on SAT-solving techniques to tackle the same problem. We refer the interested reader to~\cite{ACNS:AlbCid11,Kamal:2010,Polynomial} for more details.

%There are several papers considering cold boot attacks in the symmetric key setting. Albrecht and Cid~\cite{ACNS:AlbCid11}  focused on the recovery of symmetric encryption keys in the cold boot setting by employing polynomial system solvers, and Kamal and Youssef ~\cite{Kamal:2010} applied SAT solvers to the same problem. Further research on the development ofcold boot attacks for specific schemes can be found in~\cite{Polynomial}. Additionally,  cold boot attacks are also widely cited in the theoretically-oriented literature on leakage-resilient cryptography, but the relevance there is marginal because the cold boot attack scenario (direct access to a noisy version of the whole key) does not really apply in the leakage-resilient setting.

\subsubsection{Post-quantum setting}

Regarding the feasibility of performing this attack against post-quantum crypto-systems, several research papers have already carried out cold boot attacks on post-quantum schemes. At first, the work by Paterson et al.~\cite{10.1007/978-3-319-71667-1} explores this attack against NTRU. Their work focuses on two existing NTRU implementations, the \texttt{ntru-crypto} implementation and the \texttt{tbuktu/Bouncy Castle} Java implementation. For each in-memory format analyzed in the paper, a dedicated key-recovery algorithm is presented and tested in the bit-flipping model. One of their key-recovery algorithms may recover the private key for a small and fixed  $\alpha$ and varying $\beta$ ranging from $1\%$ up to $9\%$. A follow-up work by Villanueva-Polanco~\cite{cbbliss} expands on the previous results and presents a general key-recovery strategy via key enumeration, which is successfully applied to recover BLISS private keys. Another paper by~Villanueva-Polanco~\cite{cbaluov} adjusts the previous key recovery strategy to successfully key-recovery  LUOV private keys, exploiting the fact that a LUOV private key is derived from a $256$ bit string. Additionally, these ideas are applied to tackle the key-recovery problem for toy parameters of Rainbow and McEliece Public-Key Encryption~\cite{rvthesis}. Another recent paper~\cite{cbasike} extends these ideas to successfully key-recovery Supersingular Isogeny Key Encapsulation (SIKE) Mechanism private keys. Furthermore,  Albrecht et al.~\cite{tches-2018-29059} explore cold boot attacks on post-quantum cryptographic schemes based on the ring-and module- variants of the Learning with Errors (LWE) problem. Their work concentrates on Kyber key encapsulation mechanism (KEM) and New Hope KEM, for which they present dedicated key recovery algorithms to tackle both cases in the bit-flipping model.
\subsection{Quantum Background}

Quantum registers are qubit strings whose length determines the amount of information that they can store. In superposition, each qubit in the register is in a superposition of $\ket{0}$ and $\ket{1}$, and consequently, a register of $n$ qubits is in a superposition of all $2^n$ possible bit strings represented by $n$ ``classical'' bits.

As with single qubits, the squared absolute value of the amplitude associated with a given bit string is the probability of observing that bit string upon collapsing the register to a classical state.

\subsubsection{Quantum gates}
In classical computing, binary values, as stored in a register, pass through logic gates that, given a certain binary input, produce a certain binary output. Mathematically, classical logic gates are described as boolean functions. Quantum logic gates present a certain similarity with classical gates. When a quantum logic gate is applied to quantum registers it maps the current state to another state, transforming the state until it reaches a final state, i.e., the measured state.

There are several quantum gates each one with a specific function. In this work, we will use, 1qClifford, CNOT and Toffoli gate. For more details about gates and quantum computing see~\cite{aaronson2004improved}.

%The CNOT operates on two qubits. It flips the target qubit if and only if the control qubit is $\ket{1}$. The Toffoli gate operates on three qubits, its operation is similar to CNOT but we add one extra control bit. It flips the target qubit if and only if the two control qubits are $\ket{1}$. This gate is important because it is complete for classical reversible computation: any computation can be implemented by a circuit of Toffoli gates. Another important gate is the Hadamard gate, which operates only in one qubit but it is crucial for quantum computation. This gate maps the basis state $\ket{0}$ to $\frac{1}{\sqrt{2}}(\ket{0} + \ket{1})$ and $\ket{1}$ to $\frac{1}{\sqrt{2}}(\ket{0} - \ket{1})$, which means that we create the ``superposition'' of the basis states. The superposition can be described as the probability to measure $0$ or $1$, for more details see~\cite{lo1998introduction}.

\begin{remark}
Since all evolution in a quantum system can be described by unitary matrices and all unitary transformations are invertible, all quantum computation is reversible. For a computation to be reversible the output of the computation contains sufficient information to reconstruct the input, i.e.~no input information is erased. Unless, one 
needs to measure the state, the collapse of the state, i.e., the measurement is the only non-unitary operation in quantum computing. 
\end{remark}

\section{A framework to key recovery}\label{GAKR}

According to the results by Villanueva-Polanco~\cite{keyenum}, the key-recovery problem in the cold boot attack setting can be coped with through key-enumeration techniques. We now present the key idea from that paper.

Let us assume that $\widetilde{\texttt{k}}=\widetilde{\texttt{k}}_0\widetilde{\texttt{k}}_{1}\widetilde{\texttt{k}}_{2} \cdots \widetilde{\texttt{k}}_{ W-1}$ represent the noisy bit-string of a key of bit-length $W$ obtained via a cold boot attack. This bit string can be written as a sequence of $\mathcal{N}=W/w$ chunks, where each chunk is of length $w$ bits, i.e.
$\widetilde{\texttt{k}}=\widetilde{\texttt{K}}^0\widetilde{\texttt{K}}^{1}\widetilde{\texttt{K}}^{2} \cdots \widetilde{\texttt{K}}^{W/w-1}$ with $\widetilde{\texttt{K}}^i=\widetilde{\texttt{k}}_{i \cdot w}\widetilde{\texttt{k}}_{i \cdot w+1} \ldots \widetilde{\texttt{k}}_{(i+1)\cdot w-1}$.

Let us assume we can generate full key candidates $\texttt{c}$ for the original secret key encoding. Based on Bayes's theorem, the probability of $\texttt{c}$ 
to be the correct full key candidate given the noisy version 
$\widetilde{\texttt{k}}$ is given by   
$\mathbf{P}(\texttt{c}\lvert\widetilde{\texttt{k}})=\frac{\mathbf{P}(\widetilde{\texttt{k}}\lvert
\texttt{c})\mathbf{P}(\texttt{c})}{\mathbf{P}(\widetilde{\texttt{k}})}$. 
Thus the maximum likelihood estimation method suggests choosing $\texttt{c}$ to 
maximise $\mathbf{P}(\texttt{c} \lvert\widetilde{\texttt{k}})$. 
Note that both $ \mathbf{P}(\widetilde{\texttt{k}})$ and
$\mathbf{P}(\texttt{c})$ are constants. Thus maximising it is equivalent 
to maximise $\mathbf{P}(\widetilde{\texttt{k}}\lvert\texttt{c})=(1-\alpha)^{n_{00}}\alpha^{n_{01}}\beta^{n_{10}}(1-\beta)^{n_{11}},$ where $n_{00}$ counts the positions in which both $\texttt{c}$ and $\widetilde{\texttt{k}}$ contain a $0$ bit, $n_{01}$ counts the  positions in which $\texttt{c}$ contains a $0$ bit and $\widetilde{\texttt{k}}$ contains a $1$ bit, etc. Or equivalently, choosing $\texttt{c}$ such that maximises $\log (\mathbf{P}(\widetilde{\texttt{k}}\lvert\texttt{c}))$. Therefore each candidate can be  assigned a score, viz.  $S(\texttt{c},\widetilde{\texttt{k}}):=\log (\mathbf{P}(\widetilde{\texttt{k}}\lvert\texttt{c}))$.

 Let us assume that the full key candidates \texttt{c} are written as a sequence of chunks as for $\widetilde{\texttt{k}}$, i.e. $\texttt{c}=\texttt{C}^0|| \texttt{C}^1|| \ldots|| \texttt{C}^{\mathcal{N}-1}$, where $\texttt{C}^i$ is a $w$ bit-string, then we may also assign a score $S(\texttt{C}^i,\widetilde{\texttt{K}}^i)$ to each of the at most $2^w$ values for a chunk candidate $\texttt{C}^i$. Since ${ S(\texttt{c},\widetilde{\texttt{k}}) =\sum_{i=0}^{\mathcal{N}-1}S(\texttt{C}^i,\widetilde{\texttt{K}}^i )}$, then we can build $\mathcal{N}$ lists of chunk candidates, where each contains up to $2^w$ entries. More concretely, each list contains at most $2^w$ $2$-tuples of the form $ (score,value)$, where the first component $score$ is a real number (candidate score) and the second component ${ value}$ is a $w$-bit strings (candidate value). Now note that the original key-recovery problem reduces to a enumeration problem that consists in traversing the lists of chunk candidates to produce full key candidates $\texttt{c}$ of which total scores are obtained
by summation. The enumeration problem  has been previously studied in the side-channel
analysis literature~\cite{SAC:BKMTW15, RSA:DavWoo17,EPRINT:LMMOSS16,AC:MMOS16,AC:MOOS15,CHES:PouStaGro16, SAC:VGRS12,EC:VeyGerSta13,EPRINT:BerLanVre15,CARDIS:xin, CHES:ChoPop17,INDOCRYPT:ChoPouSta16,FSE:GGPSS15,Cardis:Pou16,Cardis:Grosso19,keyenum}, and there are many algorithms that may be useful for our key-recovery setting, in particular those enumerating full key candidates in descending order based on the score component.

After acquiring the lists of chunk candidates, one can run them into a ``search'' algorithm to find the correct key. The search can be performed by a classical or a classical-and-quantum search. In the latter, it is possible to use Grover's algorithm. However, as we will see in Section~\ref{sec:hybridattack}, the algorithm requires an oracle, and the oracle needs a quantum circuit of the underlying block cipher. In this regard, the attack becomes narrower in the direction of a specific implementation.

%\section{Rollo}
%\subsection{Key Generation algorithm}
%In this section, we describe the Rollo key generation algorithm. In particular, we focus on the one implemented in the reference implementation.

%\begin{algorithm}[H]
%\caption{Rollo Key Generation algorithm.}
%\label{alg2}
%\begin{algorithmic}
% \Function{\texttt{KeyGeneration}}{}

 %\State  \texttt{rolloII\_secretKey} skTmp;
 % \State \texttt{rolloII\_publicKey} pkTmp;
% \State  \texttt{rbc\_qre} invX;

 % \State \texttt{rbc\_qre\_init\_modulus}($\texttt{params}.N$);

 % \State  \texttt{sk\_seed}[\texttt{SEED\_BYTES]}
 % \State \texttt{randombytes}(\texttt{sk\_seed},\texttt{SEED\_BYTES});

 % \State $\texttt{skTmp} \gets \texttt{secret\_key\_from\_string}(\texttt{sk\_seed})$

  %rbc_qre_init(&invX);
 % \State $\texttt{invX} \gets \texttt{rbc\_qre\_inv}(\texttt{skTmp.x})$

  %rbc_qre_init(&(pkTmp.h));
 % \State $\texttt{pkTmp.h} \gets \texttt{rbc\_qre\_mul}( \texttt{invX}, \texttt{skTmp.y});$

  % \State \texttt{secret\_key\_to\_string}(\texttt{sk}, \texttt{sk\_seed});
 % \State \texttt{public\_key\_to\_string}(\texttt{sk} + \texttt{SEED\_BYTES}, \texttt{pkTmp});
 % \State \texttt{public\_key\_to\_string}(\texttt{pk}, \texttt{pkTmp});

% \State \Return $(\texttt{sk}, \texttt{pk})$;

% \EndFunction
%\end{algorithmic}
%\end{algorithm}
\section{Recovering secret keys via a cold boot attack}
\label{sec:recovering}
 In this section, we present our hybrid key-recovery method. We first will describe Grover's algorithm and how an attacker can use it to key-search for a block cipher and then present our key-recovery method, its general running time and costs in terms of resources.

%\subsection{Assumptions}

\subsection{Grover's algorithm}

Grover's algorithm~\cite{grover1996fast} is one of the most popular quantum algorithms among cryptographers. This algorithm provides a quadratic speedup for searching an element such as a key in a keyspace. In the following, we define the search problem:
\begin{definition}
\label{def:grover_search}
For $N = 2^n$, we are given a function $f: \{0,1\}^N \to \{0,1\}$ which assumes the value $0$ for almost all entries. The goal is to find an $x$ such that $f(x) = 1$.
\end{definition}

In the classical setting, one needs to perform $\Theta(N)$ queries for finding $x$, the number of queries varies with the randomness in the search. In the quantum setting, that is, using Grover's algorithm, one needs to perform $O(\sqrt{N})$ queries. Algorithm~\ref{alg:grover} gives a high level abstraction of Grover's algorithm.

%\def\compl#1{O\left(#1\right)}
%\begin{figure}
%		{\small
%		\centering
%		\begin{mdframed}\small
%			\underline{Grover$(f, t)$:} \\
%			\begin{tabular}{ll}
%				1.& Start with $\ket{\phi_0} = \ket{1^n}$ \\
%				2.& Apply $\textbf{H}^{\otimes n}$ \\
%				3.& Repeat $\compl{\sqrt{2^n}}$ times \\
%				4.& \qquad Phase inversion: $\mathbf{U}_f \big( \mathbf{I} \otimes \textbf{H} \big)$ \\
%				5.& \qquad Inversion about the mean: $-\mathbf{I}+ 2\mathbf{A}$ \\
%				6.& Return $x=\ket{\phi}$ with $f(x) = 1.$
%			\end{tabular}
%		\end{mdframed}
%	}
%	\caption{Grover's algorithm on a list with $n$ elements (on a high level).}
%	\label{fig:grover}
%\end{figure}

\begin{algorithm}
\footnotesize
\begin{algorithmic}
\Grover {$(f,x)$}:

\State Start with $\ket{\phi_0} = \ket{0^n}$
\State Apply $\textbf{H}^{\otimes n}$
\State \textbf{Repeat} ~$\sqrt{2^n}$ times
\State \qquad Phase inversion: $\mathbf{U}_f \big( \mathbf{I} \otimes \textbf{H} \big)$
\State \qquad Inversion about the mean: $-\mathbf{I}+ 2\mathbf{\bar{X}}$ \Comment{For more details about inversion about the mean see~\cite{yanofsky2008quantum}. }
\State \Return $x=\ket{\phi}$ with $f(x) = 1.$
\EndGrover
\end{algorithmic}
\caption{Grover's algorithm on a list with $n$ elements (on a high level).}
\label{alg:grover}
\end{algorithm}

\subsubsection{Key search for a block cipher}

Grover's algorithm can be used for searching a key in a key space. However, first the attacker needs to define the Boolean function $f$ which Grover's oracle will use it. So, a general definition can be found in~\cite{JaquesNRV20} and it is as follows:
\begin{definition}

Let $\mathcal{E}=(\texttt{E}, \texttt{D})$ be a block cipher defined over $(\mathcal{K}, \mathcal{X})$, where $\mathcal{K}=\{0,1\}^W$ and $\mathcal{X}=\{0,1\}^n$.
We denote by $\texttt{E}_{\texttt{k}}(m) \in \{0,1\}^n$ the encryption of message block $m \in \{0,1\}^n$ under key $\texttt{k}$. Given $n_p$ plaintext-ciphertext pairs $(m_i,c_i)$ with $c_i = \texttt{E}_{\texttt{k}}(m_i)$. The goal is to apply Grover's algorithm to find the unknown key $k$ by defining the function $f$ as
$$
 f(\texttt{k}) =
  \begin{cases}
   1 & \text{if } \texttt{E}_{\texttt{k}}(m_i) = c_i \text{ for all } 1 \leq i \leq n_p,\\
   0       & \text{otherwise. }
   \end{cases}
$$
\end{definition}

\subsection{Our key-recovery algorithm}
\label{sec:hybridattack}

Throughout this section, we present a key-recovery method that combines key enumeration algorithms and Grover's algorithm. The first version of this set of algorithms is introduced in~\cite{SAC:MMOS17} in the context of side-channel attacks and recently has been adjusted to be used in the cold boot attack setting on the Supersingular Isogeny Key Encapsulation (SIKE) Mechanism~\cite{cbasike}.

Here we adapt it for recovering a block cipher secret key $\texttt{sk}$ from its noisy version procured via a cold boot attack.   Let us assume that a cold boot attacker has access to a noisy version $\widetilde{\texttt{sk}}$ of a secret key $\texttt{sk} \in \mathcal{K}$ and a pair $(\texttt{m},\texttt{c}) \in \mathcal{X}\times \mathcal{X}$ such that $\texttt{E}_{\texttt{sk}}(\texttt{m})=\texttt{c}$, and has estimated the values of $\alpha$ and $\beta$. The attacker's goal is to recover \texttt{sk}.

%First, we assume the cold boot attack model discussed at length in section~\ref{CBAM}. In particular, we assume an attacker has procured $\widetilde{\texttt{sk}}$ via a cold book attack . Additionally, the attacker has access to both the parameter set $\texttt{P}$ used for generating the real key pair $(\texttt{sk},\texttt{pk})$ and the public key $\texttt{pk}$.

Recall that from our discussion in section~\ref{GAKR}, we can assign scores to each chunk candidate for a chunk by using the function $S$. Let $W$ be the length of $\widetilde{\texttt{sk}}$ in bits, $w$ be the length of a chunk in bits with $w$ dividing $W$, $\eta$ be an positive integer dividing $\mathcal{N}=W/w$ and let $\mu$ be a positive integer. Algorithm \ref{generateCandidates} creates lists of chunk candidates on inputs $\widetilde{\texttt{sk}}, W, w, \eta, \mu$. The function \texttt{toWeight} on input $s$ returns a weight (a positive integer), as suggested in ~\cite{SAC:MMOS17}. Algorithm \ref{generateCandidates} makes use of a optimal key enumeration algorithm (\texttt{OKEA})~\cite{keyenum} to get the $\mu$ most high-scoring chunk candidates for the block of chunks from $i\cdot \eta$ through $i\cdot\eta+\eta-1$, for $i=0,1,\ldots, \mathcal{N}/\eta -1$.

\begin{algorithm}[ht]
\footnotesize
\caption{creates the lists of candidates.}
\label{generateCandidates}
\begin{algorithmic}[1]
\Function{\texttt{generateCandidates}}{$\widetilde{\texttt{k}},W,w,\eta, \mu$}
\State $\mathcal{N}\gets W/w$;

\State $\Gamma \gets []$
\For{$i\gets 0~\mathrm{to}~\mathcal{N}-1$}
\State $\Pi\gets []$;
\State //Extract bits from $i \cdot w$ to $(i+1)\cdot w-1$ from $\widetilde{\texttt{k}}$
\State $\texttt{K} \gets \texttt{extract}(\widetilde{\texttt{k}}, i \cdot w,(i+1)\cdot w-1)$;
\For{$c~\in~\{0,1\}^w$}
  \State $s\gets \texttt{toWeight}(S(c,\texttt{K})$;
  \State $\Pi.\texttt{append}((s, c))$;
\EndFor
\State $\texttt{sort}(\Pi)$; //decreasing order per score.
\State $\Gamma.\texttt{append}(\Pi)$
%\State $i \gets i+1$;
\EndFor
\State $\texttt{L}=[]$;

\State $\xi \gets \mathcal{N}/\eta$;
\For{$i\gets0~\mathrm{to}~\xi-1 $}

  \State $\texttt{OKEA}.\texttt{init}(\Gamma [i \cdot \eta], \Gamma[i \cdot \eta+1],\ldots, \Gamma[i \cdot \eta+\eta-1])$;

  \State $\Pi=[]$;
  \For{$j\gets0~\mathrm{to}~\mu-1$}
   \State // $s$ is the total score of $c$.
    \State // $c$ is a bitstring of $\eta \cdot w$ bits
    \State $ (s,c) \gets \texttt{OKEA}.\texttt{getNext}()$;
    \State $\Pi.\texttt{append}((s,c))$;
    %\State $j \gets j+1$;

   \EndFor
  \State $\texttt{L}.\texttt{append}(\Pi)$;
%\State $i \gets i+1$;
\EndFor

\State \Return $\texttt{L}$;
\EndFunction
\end{algorithmic}
\end{algorithm}

Given the weights $B_1, B_2$, Algorithm~\ref{createb} constructs a two dimensional array $\texttt{B}$ with $\xi\times B_2$ entries. For
$i = \xi - 1$ and $0 \leq  b < B_2$, the entry $\texttt{B}[i][b]$ contains the number of chunk candidates such
that their total score plus $b$ lies in the interval $[B_1,B_2)$. Therefore, $\texttt{B}[i][b]$ is given by the number
of chunk candidates $L[i][j]$
, $0 \leq j < \mu$
, such that $B_1-b\leq L[i][j]
.score < B_2 - b$.

On the other hand, for $i = \xi-2, \xi-3, \ldots, 0$, and $0 \leq b < B_2$, the entry $\texttt{B}[i][b]$ contains
the number of chunk candidates that can be constructed from the chunk $i$ to the chunk
$\xi-1$ such that their total score plus $b$ lies in the interval $[B_1,B_2)$. Therefore, $\texttt{B}[i][b]$ may be
calculated as follows. For $0 \leq j < \mu$
, $\texttt{B}[i][b] =\texttt{B}[i][b]+ \texttt{B}[i+1][b+L[i][j].score]$
 if $b +L[i][j].score < B_2$. Note that By construction, $\texttt{B}[0][0]$ is the total number of full key candidates with weights in the interval $[B_1, B_2)$.

\begin{algorithm}[ht]
\footnotesize
\caption{constructs the two dimensional array $\texttt{B}$.}
\label{createb}
\begin{algorithmic}[1]
\Function{\texttt{create}}{$\texttt{L},B_1, B_2,W,w,\mu$}
\State $\mathcal{N}\gets W/w$;
\State $\xi \gets \mathcal{N}/\eta$;
\State$i \gets \xi-1$;
\State $\texttt{B} \gets [[0]*{B_2}]*{\xi}$;

\For{$b \gets 0~to~B_2-1$ }
\For{$j \gets 0~to~\mu-1$ }
\State $s \gets \texttt{L}[i][j].score$;
\If{$ B_1-b \leq s <B_2-b$ }
\State $\texttt{B}[i][b] \gets \texttt{B}[i][b]+1$;

\EndIf
\EndFor
\EndFor
\For{$i \gets \xi- 2~to~0$ }
\For{$b \gets 0~to~B_2-1$ }
\For{$j \gets 0~to~\mu -1$ }
\State $s \gets \texttt{L}[i][j].score$;
\If{$b+ s<B_2$ }
\State $\texttt{B}[i,b] \gets \texttt{B}[i][b]+\texttt{B}[i+1][b+s]$;

\EndIf
\EndFor
\EndFor
\EndFor
\State \Return $\texttt{B}$;
\EndFunction
\end{algorithmic}
\end{algorithm}

Algorithm~\ref{rank} simply constructs the matrix $\texttt{B}$ by calling \texttt{create} and then computes the total number of full key candidates with weights in $[B_1, B_2)$ by returning $\texttt{B}[0][0]$.

\begin{algorithm}[ht]
\footnotesize
%\caption{ returns the number of key candidates in a given range.}
\caption{ computes the number of full key candidates in $[B_1, B_2)$.}
\label{rank}
\begin{algorithmic}[1]
\Function{\texttt{rank}}{$\texttt{L},B_1, B_2,W,w,\eta,\mu$}
\State $\texttt{B} \gets \texttt{create}(\texttt{L},B_1, B_2,W,w, \eta,\mu)$;
\State \Return $\texttt{B}[0,0]$
\EndFunction
\end{algorithmic}
\end{algorithm}

We now present Algorithm~\ref{getKey}. This algorithm returns the full key candidate $\texttt{k}_r$ with weight in the interval $[B_1, B_2)$, with $r \in \{1, 2, 3\ldots ,\texttt{B}[0][0]\}$. By construction the output of Algorithm~\ref{getKey} is deterministic in the sense that for given fixed values of $\texttt{L},\texttt{B}, B_1, B_2,W,w,\eta,\mu$ and $r$, Algorithm~\ref{getKey}  will return the same key $\texttt{k}_r$.

Indeed, let us assume that $\texttt{L},\texttt{B}, B_1, B_2,W,w,\eta,\mu$ and $r$ are inputs to Algorithm~\ref{getKey}. We first analyse the lines from \ref{line_7} to \ref{line_20} of Algorithm~\ref{getKey}.  Let us fix $i \in \{0,\ldots, \xi-2\}$. For $j \in \{0,\ldots, \mu-1\}$, the condition of the line \ref{condi} verifies whether $r$ is less than 
the number of chunk candidates that can be constructed from the chunk $i+1$ to the chunk $\xi-1$ such that their total score plus $b+s$ lies in the interval $[B_1,B_2)$. If so, the algorithm finds the proper $j$ for the fixed $i$, then concatenate the chunk candidate $\texttt{L}[i][j].candidate$ to $\texttt{k}$ and updates $b$ as $b \gets b+ s$. Otherwise $r$ is updated as $r \gets r-\texttt{B}[i+1][b+s]$.  Similarly, the block of instructions from the line~\ref{line_21} to the line~\ref{line_29} finds the proper $j$ for $i=\xi-1$. Note that the selection of $j$'s are determined by the input parameters. Hence, for given fixed values of $\texttt{L},\texttt{B}, B_1, B_2,W,w,\eta,\mu$ and $r$, Algorithm~\ref{getKey}  will return the same key $\texttt{k}_r$.

\begin{algorithm}[ht]
\footnotesize
\caption{ returns the full key candidate $\texttt{k}_r$ with weight in the interval $[B_1, B_2).$}
\label{getKey}
\begin{algorithmic}[1]
\Function{\texttt{getKey}}{$\texttt{L}, \texttt{B}, B_1, B_2, W,w,,\eta,\mu,r$}
\State $\mathcal{N} \gets W/w$
\State $\xi \gets \mathcal{N}/\eta$

\If{$r>\texttt{B}[0][0]$}
\State \Return $\bot$
\EndIf
\State $\texttt{k}  \gets \epsilon$; //empty string \label{line_7}

\State $b \gets 0$;
\For{$i \gets 0~to~\xi-2$ }
\For{$j \gets 0~to~\mu-1$ }
\State $s \gets \texttt{L}[i][j].score$
\If{$r \leq \texttt{B}[i+1][b+s]$} \label{condi}
\State $\texttt{k} \gets \texttt{k}  \parallel \texttt{L}[i][j].candidate$;
\State $b \gets b+ s$;
\State \textbf{break} j;

\EndIf

\State $r \gets r-\texttt{B}[i+1][b+s]$;
\EndFor
\EndFor \label{line_20}

\State $i \gets \xi-1$;\label{line_21}

\For{$j \gets 0~to~\mu-1$ } 
\State $s \gets \texttt{L}[i][j].score$
\State $x \gets (B_1-b\leq s <B_2-b)?1:0$;
\If{$r \leq x$}
\State $\texttt{k} \gets \texttt{k}  \parallel \texttt{L}[i][j].candidate$;
\State \textbf{break} j;

\EndIf

\State $r \gets r-x$;
\EndFor\label{line_29}

\State \Return $\texttt{k}$
\EndFunction
\end{algorithmic}
\end{algorithm}

For completeness, we present Algorithm~\ref{keySearch} that enumerates and tests all full key candidates with weight in the interval $[B_1, B_2)$ in a classic way (without a quantum algorithm). The function $\texttt{T}$ is a boolean function that returns $1$ if $\texttt{k}$ satisfies some specific condition. Otherwise, it returns $0$.

\begin{algorithm}[ht]
\footnotesize
\caption{ enumerates and tests all full key candidates with weight in the interval $[B_1, B_2).$}
\label{keySearch}
\begin{algorithmic}[1]
\Function{\texttt{keySearch}}{$ \widetilde{\texttt{k}},B_1, B_2, W,w,\eta,\mu$}
\State $\texttt{L} \gets \texttt{generateCandidates}(\widetilde{\texttt{k}},W,w,\eta,\mu)$;
\State $\texttt{B} \gets \texttt{create}(\texttt{L},B_1, B_2,W,w,\eta,\mu)$;
\State $r \gets 1$;
\While{$True$}
\State $\texttt{k} 	\gets \texttt{getKey}(\texttt{L}, \texttt{B}, B_1, B_2, W,w,\eta,\mu,r)$;
\If{$\texttt{k}=\bot$}
\State \textbf{break};
\EndIf
\If{$\texttt{T}(\texttt{k})=1$}
\State \textbf{break};
\EndIf
\State $r \gets r+1$;
\EndWhile
\State \Return $\texttt{K}$;
\EndFunction
\end{algorithmic}
\end{algorithm}

We now present Algorithm~\ref{QKS} that performs a quantum key enumeration over an interval with roughly $e$ full key candidates. In particular, it searches over an interval of the form $[B_{min}, B_{e})$, where  $B_{min}$ is the minimum weight that a full candidate can attain given the list $\texttt{L}$ and $B_{e}$ is a calculated weight to guarantee the number of full candidates with weights in the interval $[B_{min}, B_{e})$ will be roughly $e$.  Recall that $\texttt{L}$ contains  $\xi= \mathcal{N}/\eta$ lists of chunk candidates. Therefore we can calculate the value $B_{min}$  by summing the score of the first chunk candidate of each list contained in $\texttt{L}$.

We recall that Algorithm~\ref{QKS} is ``generic'', that is, it uses Grover's algorithm in line 11 to speed up a small set of keys. The advantage of this approach is that one can attack a more broad spectrum of symmetric ciphers. 

\begin{algorithm}[ht]
\footnotesize
\caption{ performs a quantum key enumeration over a interval with roughly $e$ full key candidates.}
\label{QKS}
\begin{algorithmic}[1]
\Function{\texttt{QKS}}{$\widetilde{\texttt{k}},e, W, w,\eta,\mu$}
\State $\texttt{L} \gets \texttt{generateCandidates}(\widetilde{\texttt{k}},W,w,\eta,\mu) $;
\State $B_{min} \gets \texttt{getMinimunScore}(\texttt{L})$
\State $B_1 \gets B_{min}$;
\State $B_2 \gets B_{min}+1$;
\State $s \gets 0$;
\State  Find $B_e$ s.t. $\texttt{rank}(\texttt{L},B_1, B_e,W,w,\eta,\mu) \approx e$;
\While{$B_1\leq B_e$}
\State \label{QKS:B}$\texttt{B} \gets \texttt{create}(\texttt{L},B_1, B_2,W,w,\eta,\mu)$;
\State\label{QKS:f} $\texttt{f}(\cdot) \gets \texttt{T}(\texttt{getKey}(\texttt{L},\texttt{B}, B_1, B_2,W,w,\eta,\mu,\cdot))$;
\State \label{QKS:G} Call Grover's algorithm  with  $ \texttt{f}$

\If{a marked element $r$ is found}
\State \Return  $ \texttt{getKey}(\texttt{L},\texttt{B}, B_1, B_2,W,w,\eta,\mu, r)$;
\EndIf

\State $s \gets s+1$;
\State $B_1 \gets B_{2}$;
\State  Find $B_2$ s.t. $\texttt{rank}(\texttt{L},B_1, B_2,W,w,\eta,\mu) \approx 2^s$
\EndWhile
\State \Return $\bot$;
\EndFunction
\end{algorithmic}
\end{algorithm}

\subsubsection{Quantum circuit for \texttt{f}}

The quantum circuit for \texttt{f} (Line \ref{QKS:f} of Algorithm \ref{QKS}) can be seen as the oracle implementation of $\texttt{E}$. In particular, given a plain-text/cipher-text pair $(\texttt{m}, \texttt{c})$, \texttt{T} is defined as

$$
 \texttt{T}(\texttt{k}) =
  \begin{cases}
   1 & \text{if } \texttt{E}_{\texttt{k}}(\texttt{m}) = \texttt{c}\\
   0       & \text{otherwise. }
   \end{cases}
$$

\noindent where $\texttt{k}=\texttt{getKey}(\texttt{L},\texttt{B}, B_1, B_2,W,w,\eta,\mu,r)$ and $r \in \{1, 2, 3\ldots ,\texttt{B}[0][0]\}$. That is, Grover's algorithm is run to search a key in the space $\mathcal{K}_1$ generated by \texttt{getKey} for fixed values of $\texttt{L},\texttt{B}, B_1, B_2,W,w,\eta,\mu$ and $r \in \{1, 2, 3\ldots ,\texttt{B}[0][0]\}$. In this regard, each attempt for running Grover's algorithm with oracle \texttt{f} will cost $O(\sqrt{\texttt{B}[0][0]} \approx 2^{s/2})$, where $s=0,1,2, \ldots,.$

In a practical example, let us suppose that Algorithm~\ref{QKS} at line \ref{QKS:B} generates a matrix  \texttt{B} such that  $\texttt{B}[0][0]= 2^s$, where $s = 16$. Therefore, $2^{16}$ candidates need to be tested. At line \ref{QKS:G}, Grover's algorithm is run with an oracle $\texttt{f}$, which can be constructed from the result by~\cite{JaquesNRV20}, to check if it can find the correct answer. Given we have an unique result, we will need to run this algorithm $O(\sqrt{2^{16}}=2^8)$ times, until we have reached our correct solution or not.

As pointed out, a critical component of our algorithm is the quantum oracle, so we will next present how to implement the quantum oracle for several block ciphers, namely, AES, PRESENT, and GIFT. Afterward, in Section~\ref{sec:Picnic}, we further evaluate our algorithm for LowMC, in particular in the context of Picnic, the post-quantum signature algorithm currently being assessed by the NIST standardization process.

\subsubsection{Quantum AES}
As previously mentioned, quantum computations need to be reversible. Also, the oracle $\mathcal{O}$ present in Grover's algorithm implements the block cipher as a reversible function. In~\cite{Grassl2016}, the authors give the first version of a reversible AES. Their seminal work generate other implementations in the literature such as~\cite{Almazrooie2018, Kim2018, langenberg2020reducing, JaquesNRV20, DavenportP20}.

AES is a block cipher, designed by Daemen and Rijmen~\cite{daemen02}. It is based on Rijndael but only provides $128$-bit blocks. AES has different transformations operating on an intermediate result that is called \texttt{State}. The state can be seen as an array of bytes, with four rows and four columns. The number of rounds $N_r$ depends on the size of the key, e.g., AES-$128$ performs $10$ rounds, AES-$192$ performs $12$ rounds and AES-$256$ performs $14$ rounds.

In the encryption process with AES, one needs first to perform key addition, denoted by \texttt{AddRoundKey}, followed by $N_r-1$ executions of \texttt{Round}, and finally one application of \texttt{FinalRound}. The \texttt{Round} function is the application of $4$ transformations which are \texttt{SubBytes}, \texttt{ShiftRows}, \texttt{MixColumns} and \texttt{AddRoundKey}. The \texttt{FinalRound} consists of the application of \texttt{SubBytes}, \texttt{ShiftRows} and \texttt{AddRoundKey}.  Algorithm~\ref{AESHL} shows, in a pseudo C language, how those rounds are put together. One advantage of AES is that one just needs to implement the transformation functions and then reuse them in the rounds.
\begin{algorithm}[ht]
\caption{High level description of AES.}
\label{AESHL}
\begin{algorithmic}
\Function{\texttt{AES}}{$\texttt{State},\texttt{CipherKey}$}
  \State \texttt{KeyExpansion}(\texttt{CipherKey}, \texttt{ExpandedKey});
   \State \texttt{AddRoundKey}(\texttt{State}, \texttt{ExpandedKey}[0]);
   \For{$(i\gets 1,i<N_r, i\gets i+1)$}
    \State \texttt{Round}(\texttt{State}, \texttt{ExpandedKey}$[i]$);
   \EndFor
   \State \texttt{FinalRound}(\texttt{State}, \texttt{ExpandedKey}$[N_r]$);
\EndFunction
\end{algorithmic}
\end{algorithm}

%\begin{lstlisting}[language=C,escapeinside={(*}{*)}, caption={High level description of AES.}, label={lst:aes_high_pseudo}]
%AES(State, CipherKey){
%  KeyExpansion(CipherKey, (*$ExpandedKey$*));
%  AddRoundKey(State, (*$ExpandedKey$*)[0]);
%  for(i = 1; i < (*$N_r$*); i++){
%    Round(State, (*$ExpandedKey$*)[i]);
%  }
%  FinalRound(State, (*$ExpandedKey$*)[(*$N_r$*)]);
%}
%\end{lstlisting}

In the latest literature, we can see an improvement in the quantum circuit developed to AES. In our case, we will consider the implementation in~\cite{JaquesNRV20} since it gives the lowest depth. We consider the ``in-place'' setting, more details in~\cite[Sec. 4.6]{JaquesNRV20}. Table~\ref{tab:costAES} gives the number of gates necessary to run AES in Grover's algorithm.

\begin{table}[ht]
\caption{Number of quantum gates for the full encryption circuit for AES presented in~\cite[Sec. 4.6]{JaquesNRV20}.}
\label{tab:costAES}
\begin{center}
\begin{tabular}{c|rrr}
\toprule
AES & CNOT    & 1qCliff & T     \\
\midrule
AES-$128$          & $291\,150$  & $83\,116$    & $54\,400$  \\
AES-$192$          & $328\,612$ & $93\,160$    & $60\,928$ \\
AES-$256$         & $402\,878$ & $114\,778$   & $75\,072$ \\
\bottomrule
\end{tabular}
\end{center}
\end{table}

\subsubsection{Quantum PRESENT \& Quantum GIFT}

PRESENT~\cite{YangZSAG15} and GIFT~\cite{BanikPPSST17} follow the 
block cipher construction, that is, both schemes have a
certain number of rounds in which they apply an Sbox transformation 
followed by a permutation. However, each of them has some difference. 
For PRESENT, the first operation is the addition of the round key, while, for GIFT, the first operation is the Sbox transformation. 

PRESENT has block sizes of $64$ bits, and GIFT uses $64$ and $128$ bits blocks. PRESENT support $80$-bit key size, and both of them support $128$-bit key size. 
More details can be found in the original papers~\cite{YangZSAG15, BanikPPSST17}.

Fortunately, there are implementations of both of them in the quantum world, that is, there are 
reversible implementations using quantum gates. The work in~\cite{app11114776} provides 
a deeper analysis of the quantum circuit. Table~\ref{tab:costsPG} show the number of 
gates for PRESENT and GIFT. The authors in~\cite{app11114776} give the estimation using CNOT and Toffoli gates, 
in order to use in our work we use the same decomposition as~\cite{Grassl2016} and decompose $1$ Toffoli gate as $7$ T gates + $8$ Clifford gates. We remark that this gives an upper bound on the number of T gates as we use the generic decomposition; the circuits above could be built using T-gates directly and possibly use fewer T gates~\cite{PhysRevA.70.052328}.

\begin{table}[ht]
\caption{Number of quantum gates for the full encryption circuit for PRESENT and GIFT presented in~\cite{app11114776}.}
\label{tab:costsPG}
\begin{center}
\begin{tabular}{c|rrr}
\toprule
Block cipher & CNOT    & 1qCliff & T     \\
\midrule
PRESENT-$64$/$80$          & $18\,892$  & $67\,456$    & $59\,024$  \\
PRESENT-$64$/$128$           & $19\,608$ & $71\,424$    & $62\,496$ \\
GIFT-$64$/$128$           & $7\,424$ & $57\,344$    & $50\,176$ \\
GIFT-$128$/$128$           & $12\,288$ & $98\,304$    & $86\,016$ \\
\bottomrule
\end{tabular}
\end{center}
\end{table}

\paragraph{Generic Implementation and Different ciphers.} We present the costs to implement AES, PRESENT and GIFT into a quantum computer. As mentioned before, our attack is generic, and one can easily replace the function $\texttt{f}(\cdot)$ in Algorithm~\ref{QKS} by one of those implementations. In the following, we will focus in LowMC given that it is the one used in Picnic, which is the scope of this work.

\section{Cold boot attacks on Picnic}
\label{sec:Picnic}

In this section, we further evaluate our algorithm for LowMC, in particular in the context of Picnic, the post-quantum signature algorithm currently being assessed by the NIST standardization process. We first describe the key-generation algorithm as it is implemented in ~\cite{pinicRI}. We then describe the inner workings of LowMC and its Quantum version, and then the costs and success rate of our algorithm in this context.

\subsection{Picnic key generation algorithm}

In our analysis, we use the current reference implementation of Picnic~\cite{pinicRI}. Algorithm~\ref{alg:keygen} summarizes the process of key generation.
%We present the key generation algorithm for Picnic as it is currently implemented in Picnic's reference implementation~\cite{pinicRI}.

\begin{algorithm}[ht]
\footnotesize
\caption{Picnic's Key Generation Algorithm}
\label{alg:keygen}
\begin{algorithmic}[1]
\Function{\texttt{keygen}}{\texttt{P}}
    \State $\texttt{sk} \gets \texttt{randBytes(P.stateSizeBytes)};$
    \State $\texttt{zeroTrailBits}(\texttt{sk}, \texttt{P.stateSizeBits})$;
    \State $\texttt{m} \gets \texttt{randBytes}(\texttt{P.stateSizeBytes});$
    \State $\texttt{zeroTrailBits}(\texttt{m}, \texttt{P.stateSizeBits})$;
    \State $\texttt{c} \gets \texttt{LowMCEnc}(\texttt{m},\texttt{sk},\texttt{P} )$
    \State $\texttt{pk} \gets (\texttt{m},\texttt{c});$
    \State \Return $\texttt{ sk}, \texttt{pk};$
\EndFunction
\end{algorithmic}
\end{algorithm}
As one can see, the input of the function \texttt{KeyGen} is \texttt{P}, which represents an instance of a structure to store a parameter set (\texttt{paramset\_t}). This structure points to a relatively big set of fields. In particular, the field $\texttt{stateSizeBytes}$ refers to the number of bytes needed to store $\texttt{stateSizeBits}$ bits, which is the bit length of $\texttt{sk},\texttt{m}$ and $\texttt{c}$. In particular, Table~\ref{tab:my-table} shows the values of both \texttt{stateSizeBits} and \texttt{stateSizeBytes} for each Parameter Set for Picnic, as defined in the Picnic reference implementation file \texttt{picnic.c}~\cite{pinicRI}.

%\begin{table*}[ht!]
%\caption{Values of both \texttt{stateSizeBits} and \texttt{stateSizeBytes} for each Parameter Set for Picnic}
%\label{tab:my-table}
%\begin{center}
%\resizebox{\textwidth}{!}{
%\begin{tabular}{|c|c|c|}
%\hline
%Parameter Set  & \texttt{stateSizeBits} & \texttt{stateSizeBytes} \\ \hline
%\texttt{picnic-L1-FS}   &   $128$       &     $16$           \\ \hline
%\texttt{picnic-L1-UR}   &   $128$       &     $16$           \\ \hline
%\texttt{picnic-L1-full} &   $129$       &     $17$           \\ \hline
%\texttt{picnic3-L1}     &   $129$       &     $17$           \\ \hline
%\texttt{picnic-L3-FS}   &   $192$       &     $24$             \\ \hline
%%\texttt{picnic-L3-UR}   &   $192$       &     $24$          \\ \hline
%\texttt{picnic-L3-full} &   $192$       &     $24$           \\ \hline
%\texttt{picnic3-L3}     &   $192$       &     $24$           \\ \hline
%\texttt{picnic-L5-FS}   &   $256$       &     $32$           \\ \hline
%\texttt{picnic-L5-UR }  &   $256$       &     $32$          \\ \hline
%\texttt{picnic-L5-full} &   $255$       &     $32$              \\ \hline
%%\texttt{picnic3-L5}     &   $255$       &     $32$           \\ \hline
%\end{tabular}}
%\end{center}
%\end{table*}

\begin{table}[ht!]
\caption{Values of both \texttt{stateSizeBits} and \texttt{stateSizeBytes} for each Parameter Set for Picnic}
\label{tab:my-table}

\begin{center}

\begin{tabular}{l|rr}
\toprule
Parameter Set  & \texttt{stateSizeBits} & \texttt{stateSizeBytes} \\
\midrule
\texttt{picnic-L1-FS}   &   $128$       &     $16$           \\
\texttt{picnic-L1-UR}   &   $128$       &     $16$           \\
\texttt{picnic-L1-full} &   $129$       &     $17$           \\
\texttt{picnic3-L1}     &   $129$       &     $17$           \\
\texttt{picnic-L3-FS}   &   $192$       &     $24$             \\
\texttt{picnic-L3-UR}   &   $192$       &     $24$          \\
\texttt{picnic-L3-full} &   $192$       &     $24$           \\
\texttt{picnic3-L3}     &   $192$       &     $24$           \\
\texttt{picnic-L5-FS}   &   $256$       &     $32$           \\
\texttt{picnic-L5-UR }  &   $256$       &     $32$          \\
\texttt{picnic-L5-full} &   $255$       &     $32$              \\
\texttt{picnic3-L5}     &   $255$       &     $32$           \\
\bottomrule
\end{tabular}

\end{center}

\end{table}

For the sake of completeness, the call to $\texttt{randBytes}(\texttt{size})$ returns a random byte array of length $\texttt{size}$, while the call to $\texttt{zeroTrailBits}(\texttt{byteArray},\texttt{bitLength})$ sets to $0$ all bits of $\texttt{byteArray}$ at position $i$ for all $\texttt{bitLength}<i\leq 8\cdot l$, where $l$ is the number of entries of $\texttt{byteArray}$. At line $6$, we see a call to LowMCEnc, the LowMC encryption algorithm, which we will describe next.

\subsection{LowMC block cipher}
LowMC~\cite{lowMC15,lowMC16} is a block cipher that tries to reduce the multiplicative complexity of circuits. Different from other block ciphers, the instantiation of LowMC is not fixed, and it depends on the choice of certain parameters such as the block size, number of S-Boxes per round, and security expectations. Besides encryption and decryption, LowMC is also a component of the Picnic signature scheme.

%Even not following the other block ciphers regarding the instantiation, LowMC encryption procedure follows the other block ciphers.
First, LowMC performs a key-whitening and then iterates a round function by $R$ times, where $R$ depends on the parameters. The round function consists of $4$ steps and is summarized as follows.
\begin{enumerate}
    \item SBoxLayer: A $3$-bit S-Box is applied to the first $3m$ bits of the state in parallel, while an identity map is applied to the remaining bits;
    \item MatrixMul: A regular matrix $L_i \in \mathbb{F}^{n\times n}_{2}$ is generated at random and the $n$-bit state is multiplied by $L_i$;
    \item ConstantAddition: An $n$-bit constant $C_i \in \mathbb{F}^{n}_{2}$ is randomly generated and then compute the addition of $n$-bit state and $C_i$;
    \item KeyAddition: A full-rank matrix $M_{i+1} \in \mathbb{F}^{n \times k}_{2}$ is randomly generated. The $n$-bit round key $K_{i+1}$ is obtained by multiplying the $k$-bit master key with $M_{i+1}$. Then, the $n$-bit state is added with $K_{i+1}$, where addition means XOR operation.
\end{enumerate}

To use LowMC in Picnic, the authors in~\cite{pinicRI} defined three levels: L1, L3, L5. For details about the construction given the parameters, we refer to the documentation in~\cite{pinicRI}.
% Gustavo I am about to send it. Are you making other changes to it?

%\section{Hybrid attack approach to LowMC}
%\label{sec:hybridattack}

\iffalse
\fi

\subsubsection{Quantum LowMC}
In this context, we will need a quantum version of LowMC. Fortunately, \cite{JaquesNRV20} presents a quantum version of LowMC with low depth in their circuit. Furthermore, the authors provide a Q\# implementation of the LowMC. We will reuse their results since it deals with the problems of building quantum circuits. Table~\ref{tab:costslowmc} shows the number of quantum gates necessary for applying the LowMC encryption. The levels L1, L3, and L5 are the security levels required by Picnic scheme.

\begin{table}[ht]
\caption{Number of quantum gates for the full encryption circuit for LowMC presented in~\cite[Sec. 5.4]{JaquesNRV20}.}
\label{tab:costslowmc}
\begin{center}
\begin{tabular}{c|rrr}
\toprule
LowMC Level & CNOT    & 1qCliff & T     \\
\midrule
L1          & $689\,944$  & $4\,932$    & $8\,400$  \\
L3          & $2\,271\,870$ & $9\,398$    & $12\,600$ \\
L5          & $5\,070\,324$ & $14\,274$   & $15\,960$ \\
\bottomrule
\end{tabular}
\end{center}
\end{table}

\begin{figure}
  \begin{center}
\includegraphics[width=0.9\columnwidth]{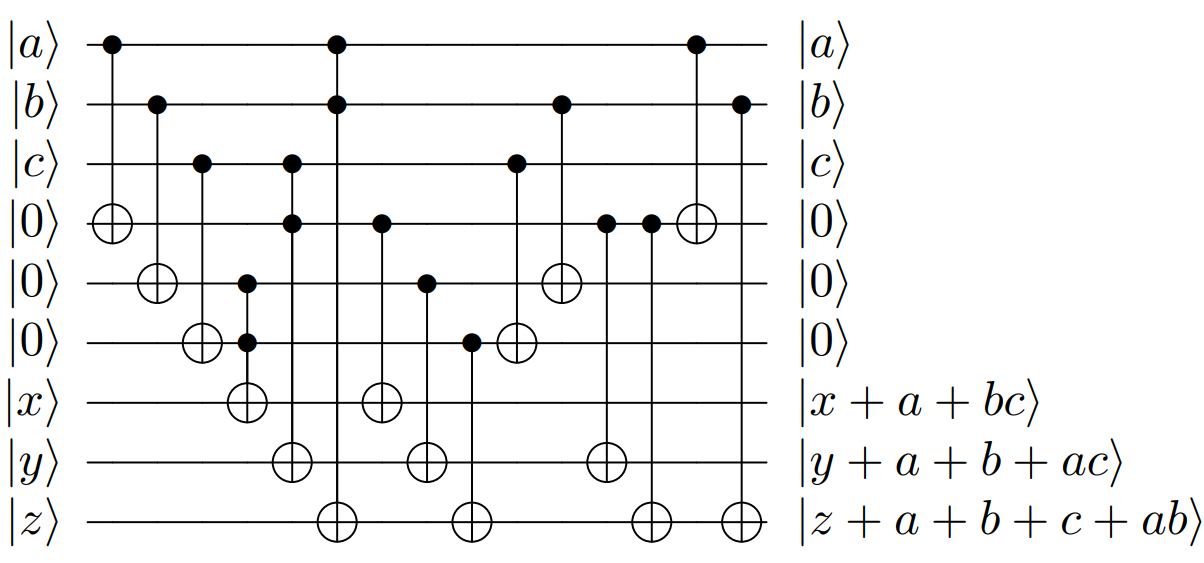}
\caption{Quantum circuit for computation of one S-Box from LowMC.}
\label{fig:lowmcs}
  \end{center}
\end{figure}

Figure~\ref{fig:lowmcs} shows the implementation of one S-Box, it is possible to notice that it requires $3$ ancillas for storing intermediate results and it requires $12$ CNOT gates and $3$ Toffoli gates. In the Picnic specification it defines that a full S-boxLayer consists of 10 parallel S-Boxes.

The AffineLayer since it is an affine transformation, it consists of a matrix multiplication following by an addition of a constant vector. The details can be seen in~\cite[Sec. 5.2]{JaquesNRV20}. The last function to describe, that is, the KeyExpansion and KeyAddition are only CNOT gates in parallel to perform the addition.

\subsection{Costs for running our key recovery algorithm}

The costs in terms of gates for running LowMC are similar to those provided in~\cite{JaquesNRV20}. The only difference for our case is that we will search in a smaller keyspace, that is, the candidates that Algorithm~\ref{QKS} generates in line 9. Table~\ref{tab:costsGrover}
shows the costs for running Grover's algorithm with the oracle provided in~\cite{JaquesNRV20}. Furthermore, we select $3$ different sizes of windows for the interval $[B_{min}, B_e)$, namely $e \in \{2^{30}, 2^{40}, 2^{50}\}$ full candidates.

\begin{table}[ht]
\caption{Total number of gates for running Grover's algorithm against LowMC.}
\label{tab:costsGrover}
\begin{center}
\resizebox{\columnwidth}{!}{
\begin{tabular}{c|rrrr}
\toprule
Value of \textbf{e} & Level & CNOT    & 1qCliff & T     \\
\midrule
\multirow{3}{*}{$30$} & L1         & $1.78 \times 10^{10}$  & $1.1 \times 10^8$    &                                         $2.16\times 10^8$  \\
                    & L3            & $5.85 \times 10^{10}$  & $2.42 \times 10^8$    & $3.24\times 10^8$  \\
                    & L5            & $1.3 \times 10^{11}$  & $3.67 \times 10^8$    & $4.11\times 10^8$  \\
\midrule
\multirow{3}{*}{$40$} & L1         & $5.68 \times 10^{11}$  & $3.24 \times 10^9$    &                                         $6.9\times 10^9$  \\
                    & L3            & $1.87 \times 10^{12}$  & $7.74 \times 10^9$    & $1.04\times 10^{10}$  \\
                    & L5            & $4.18 \times 10^{12}$  & $1.18 \times 10^{10}$    & $1.31\times 10^{10}$  \\
\midrule
\multirow{3}{*}{$50$}   & L1       & $1.82 \times 10^{13}$  & $1.04 \times 10^{11}$    &                                         $2.21\times 10^{11}$  \\
                        & L3        & $5.99 \times 10^{13}$  & $2.48 \times 10^{11}$    & $3.32\times 10^{11}$  \\
                        & L5        & $1.34 \times 10^{14}$  & $3.76 \times 10^{11}$    & $4.21\times 10^{11}$  \\
 \bottomrule
\end{tabular}
}
\end{center}

\end{table}

In our analysis, we need to consider the costs to run $O(N)$ times, since the costs provided in~\cite{JaquesNRV20} are only for $1$ query. In our case, our costs are $O(N) \times \#CNOT$,  $O(N) \times \#1qCliff$,  $O(N) \times \#T$, for CNOT, 1qCliff and T gates respectively, where $O(N)$ is taken as $\frac{\pi}{4}\sqrt{2^e}$.

\begin{remark}
It is possible to run our algorithm in parallel or reuse the circuit. Since we fix the size of window, one can pre-compute the sub-intervals  $[B_0,B_1), [B_1,B_2),\ldots, [B_{j}, B_{e})$,
where each has size $2^s$, for $s=0,1,\ldots$. One can reuse the circuit to run each chunk in sequence or run several instances of Grover's algorithm each one with their chunk of keys.
\end{remark}

\begin{remark}
Our Algorithm~\ref{QKS} is a ``hybrid'' algorithm. In our case, we are considering that everything before the Grover's call is ``classical'' computation. The same after the call, that is, when we check if the element is found. Hence, we do not need to take into account the costs of the other functions in a quantum computer besides the one in line 11. Furthermore, in our costs, we assume that we always find the element in our search. However, as a future work, we suggest that it is necessary to evaluate the other case, i.e., when the Algorithm~\ref{QKS} will run more than one time. 
\end{remark}

\subsection{Success rate of our key recovery algorithm}

In this section, we present the success rate of our key-recovery algorithm for each set of parameters defined for Picnic in~\cite{pinicRI}. The success rates are estimated by performing simulations of our key recovery algorithm for several selected hyper-parameters.

\begin{figure*}[ht!]
     \centering
     \begin{adjustbox}{minipage=\textwidth}
     \begin{subfigure}{0.45\textwidth}
         \centering
         \includegraphics[width=\textwidth]{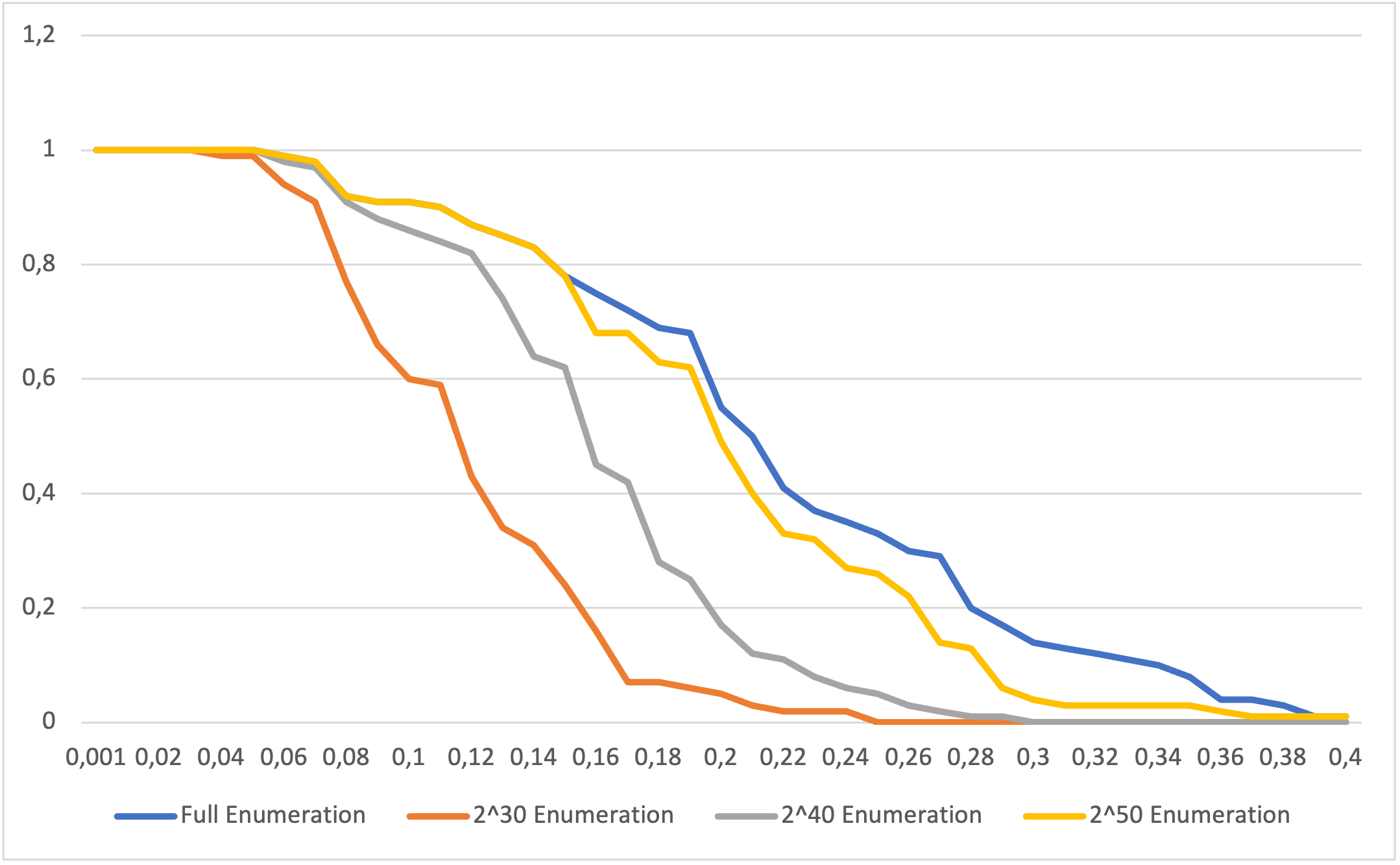}
         \caption{$\mu=256$}
         \label{fig:Picnic12_256}
     \end{subfigure}
     \begin{subfigure}{0.45\textwidth}
         \centering
         \includegraphics[width=\textwidth]{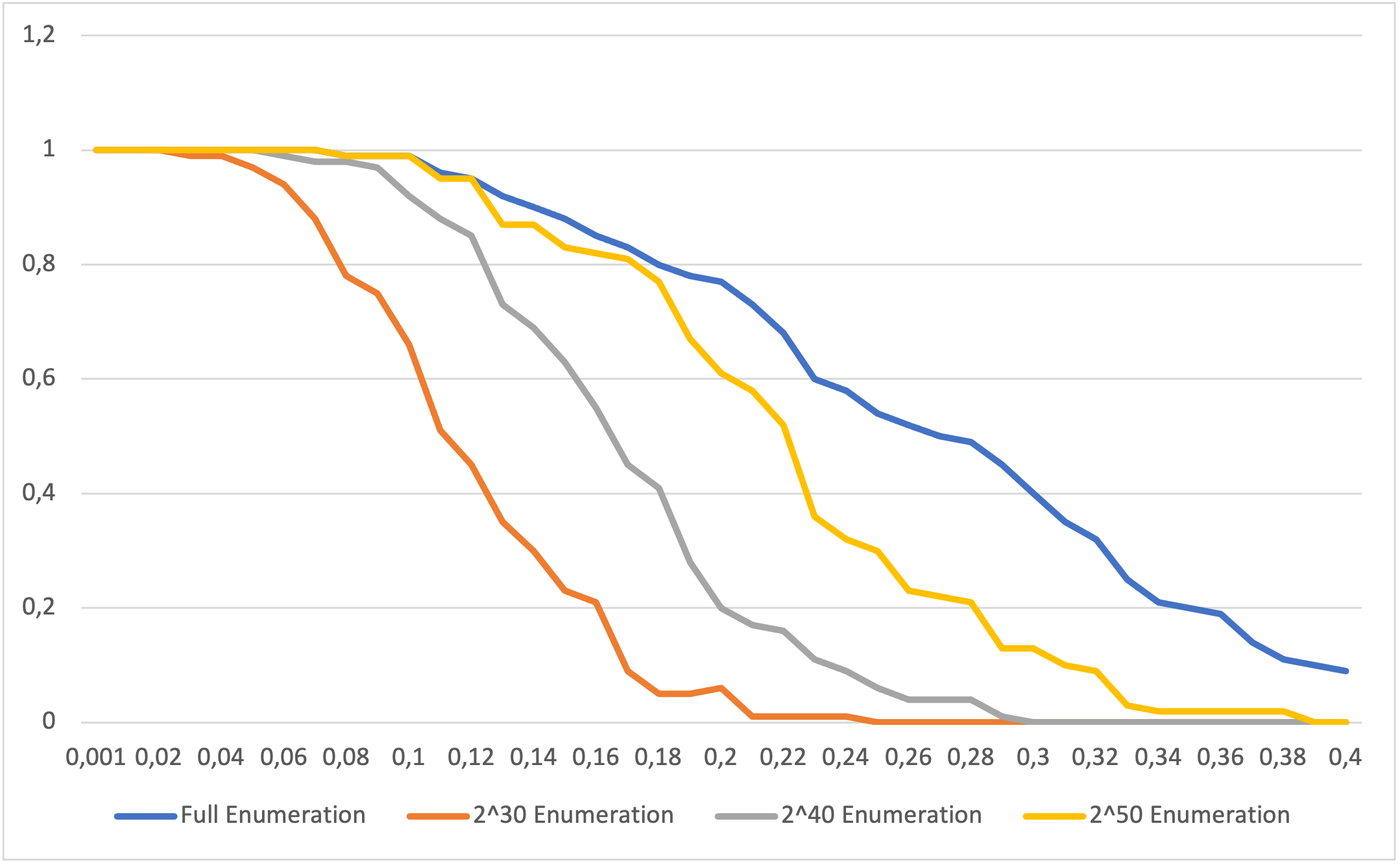}
         \caption{$\mu=512$}
         \label{fig:Picnic12_512}
     \end{subfigure}
      \begin{subfigure}{0.45\textwidth}
         \centering
         \includegraphics[width=\textwidth]{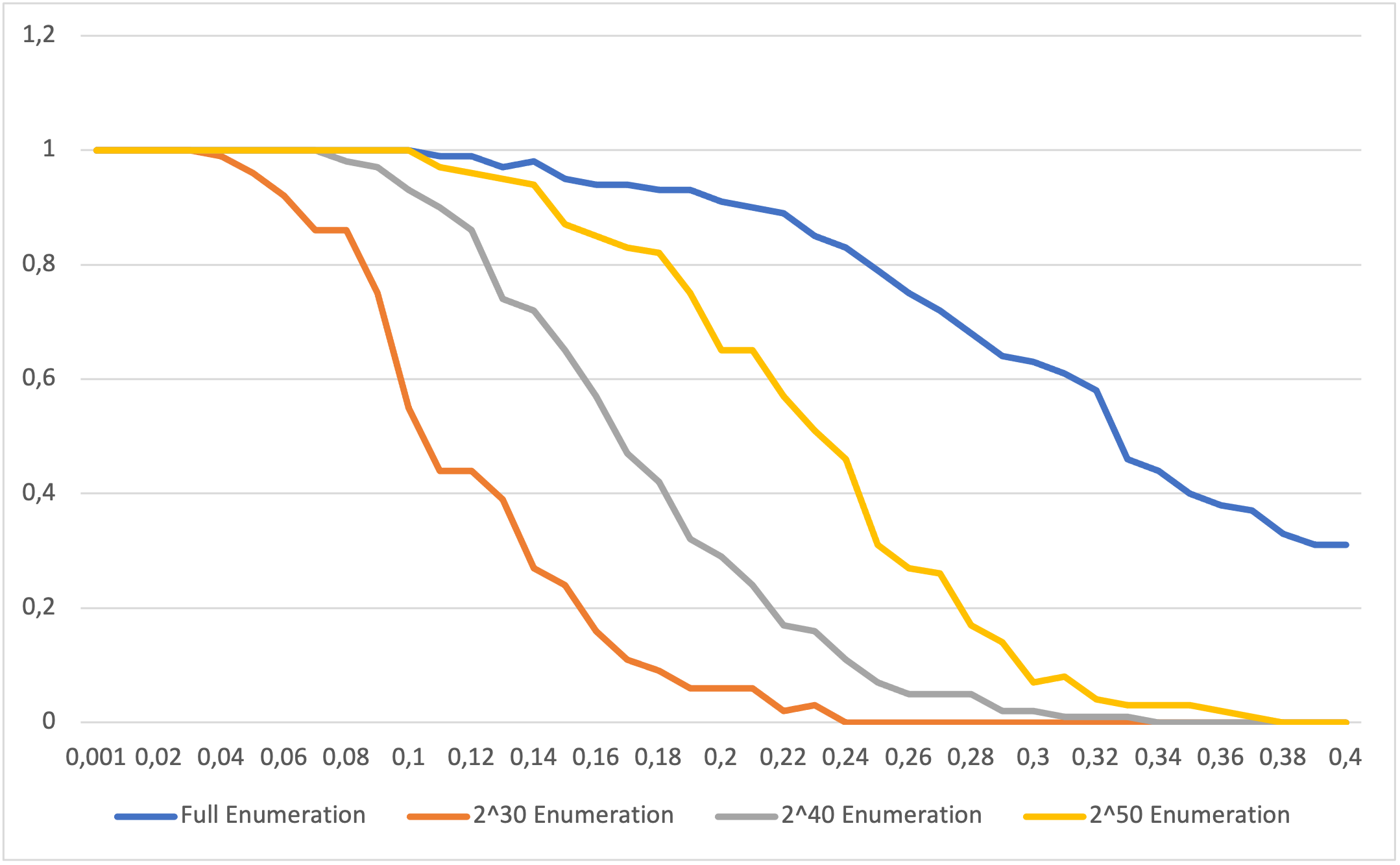}
         \caption{$\mu=1024$}
         \label{fig:Picnic12_1024}
     \end{subfigure}
     \hfill
     \begin{subfigure}{0.45\textwidth}
         \centering
         \includegraphics[width=\textwidth]{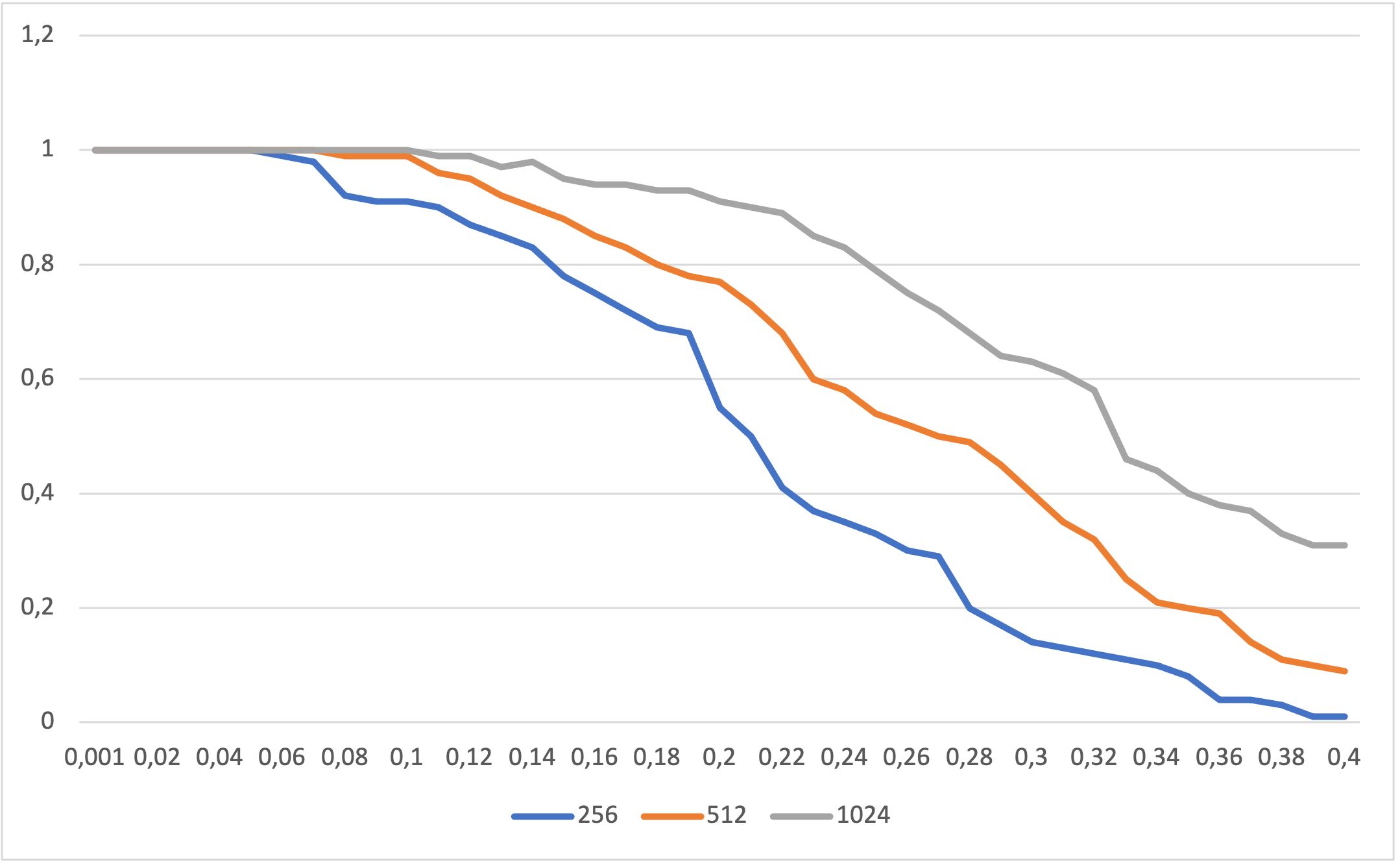}
         \caption{Full Enumeration for $\mu \in \{256,512,1024\}$}
         \label{fig:Picnic12_full}
     \end{subfigure}
        \caption{Success rate of our key recovery algorithm with $W=128, w=8, \eta=2$, $\alpha=0.001$ and $\beta \in
        \{0.001, 0.01, 0.02, \ldots, 0.4\}$ for Picnic parameters \texttt{picnic-}\{\texttt{L1-FS}, \texttt{L1-UR}, \texttt{L1-full}\} and \texttt{picnic3-L1}. The $x$-axis represents $\beta$, while $y$-axis represents the success rate. }
        \label{fig:Picnic12}
       \end{adjustbox}
\end{figure*}

We note that our key-recovery method might find $sk$ from $\widetilde{sk}$, only if each list from the list $\texttt{L}$ returned by Algorithm~\ref{generateCandidates} contains the proper chunk candidates to reconstruct $\texttt{sk}$. In such a case, a full enumeration of all candidates constructed from the lists of chunk candidates contained in $\texttt{L}$ will find the real private key.

\begin{figure}[ht!]
     \centering
     \begin{adjustbox}{minipage=\columnwidth}
     \begin{subfigure}{0.45\textwidth}
         \centering
         \includegraphics[width=\textwidth]{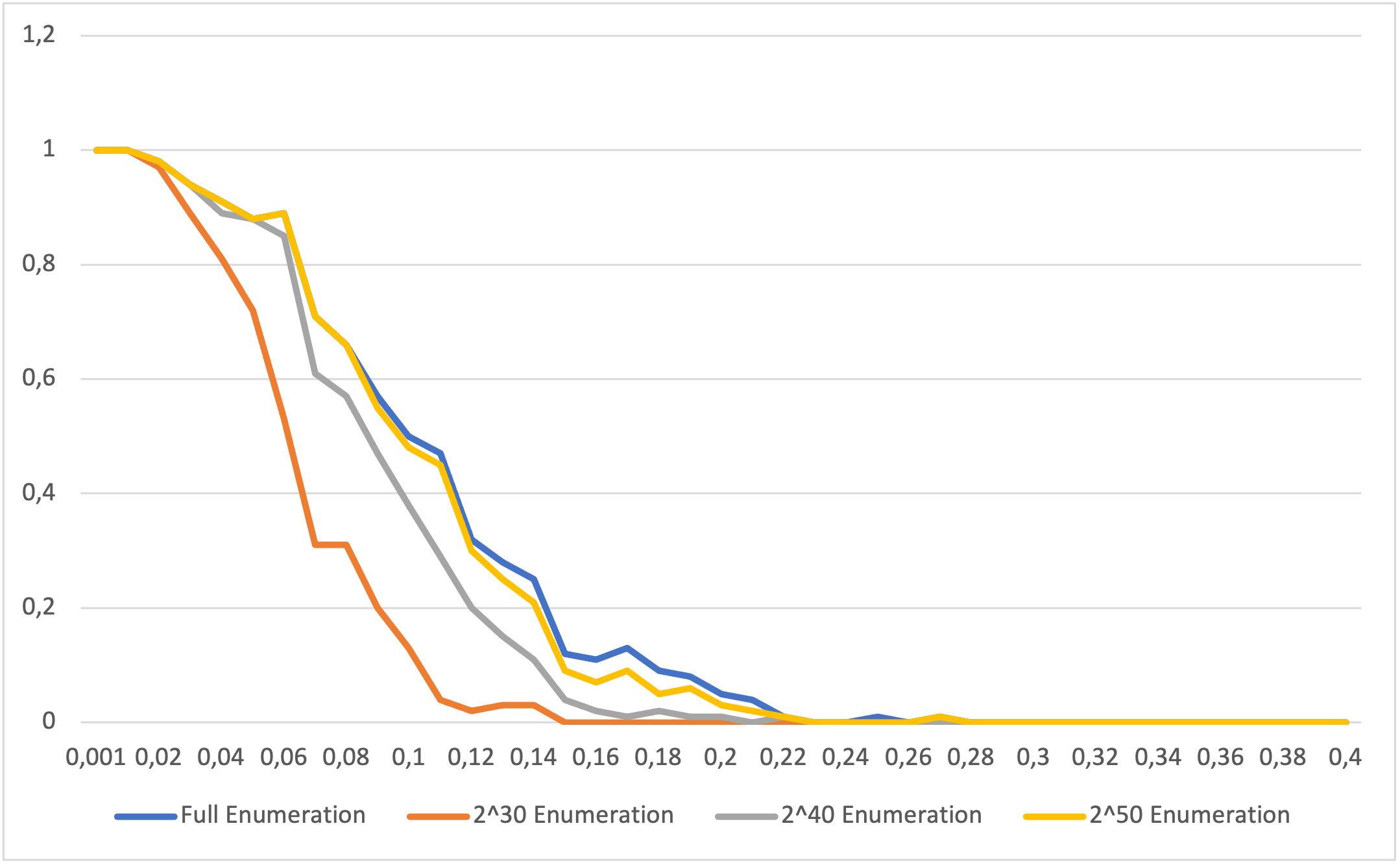}
         \caption{$\mu=256$}
         \label{fig:Picnic34_256}
     \end{subfigure}
     \hfill
     \begin{subfigure}{0.45\textwidth}
         \centering
         \includegraphics[width=\textwidth]{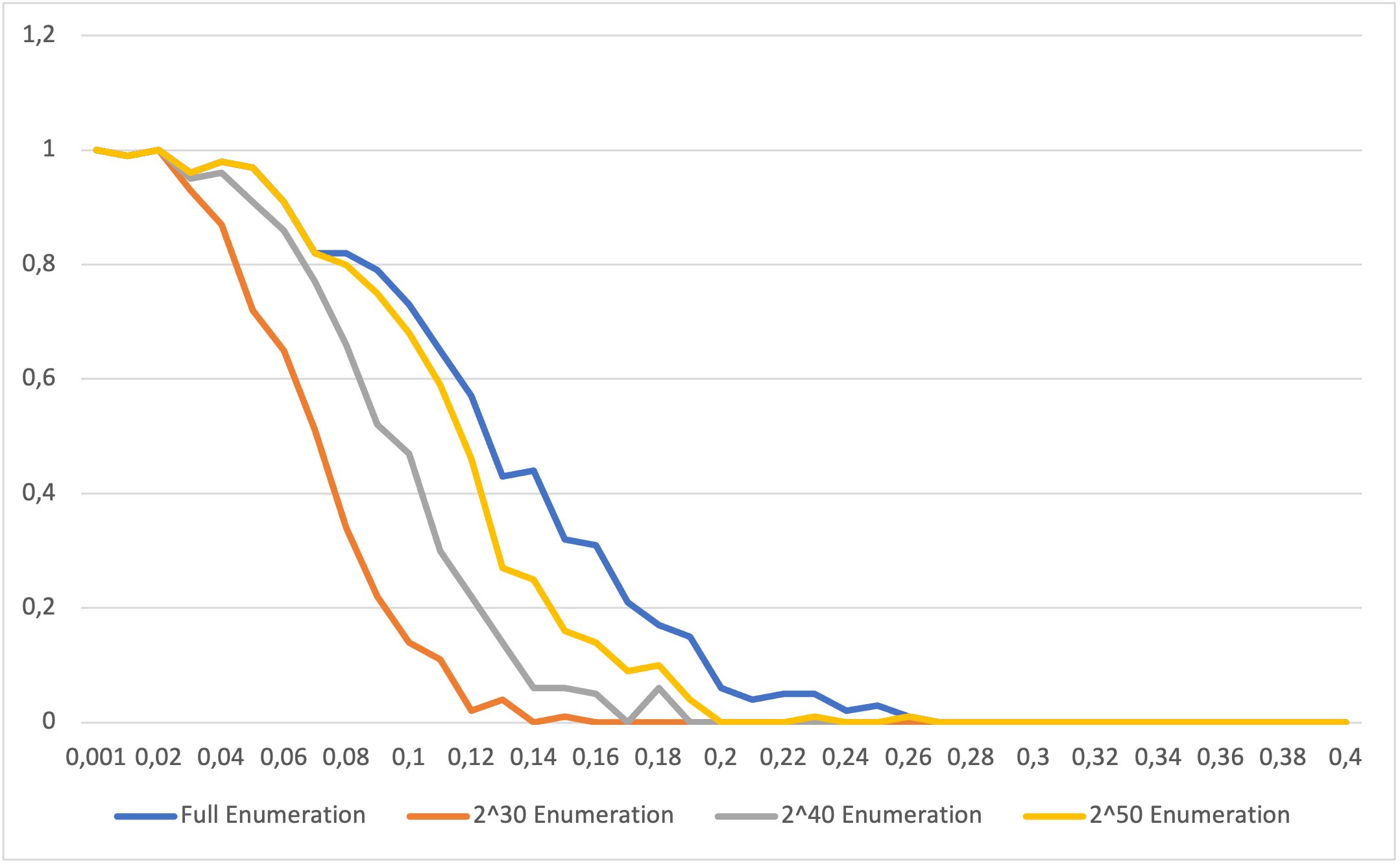}
         \caption{$\mu=512$}
         \label{fig:Picnic34_512}
     \end{subfigure}
     \begin{subfigure}{0.45\textwidth}
         \centering
         \includegraphics[width=\textwidth]{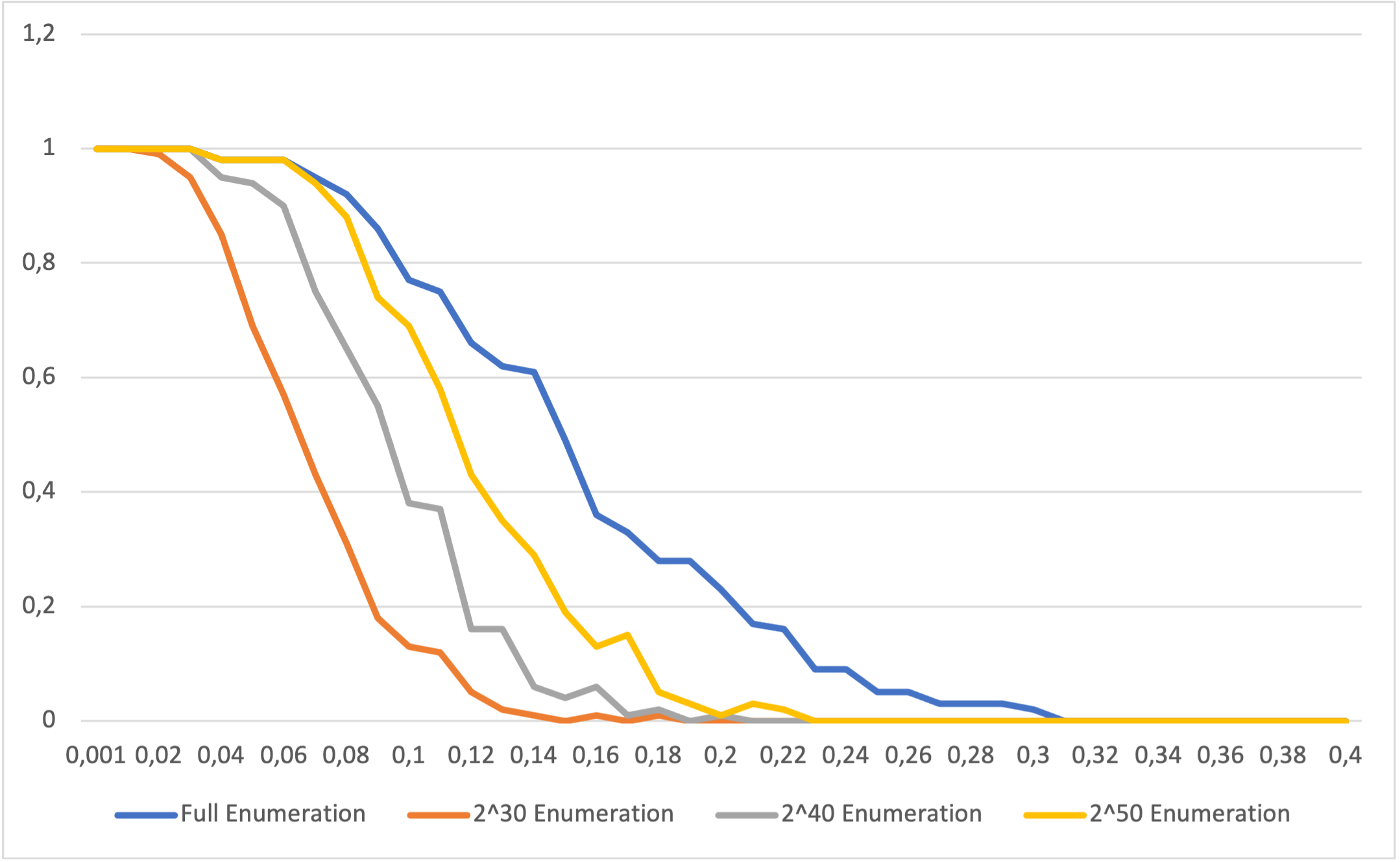}
         \caption{$\mu=1024$}
         \label{fig:Picnic34_1024}
     \end{subfigure}
     \hfill
     \begin{subfigure}{0.45\textwidth}
         \centering
         \includegraphics[width=\textwidth]{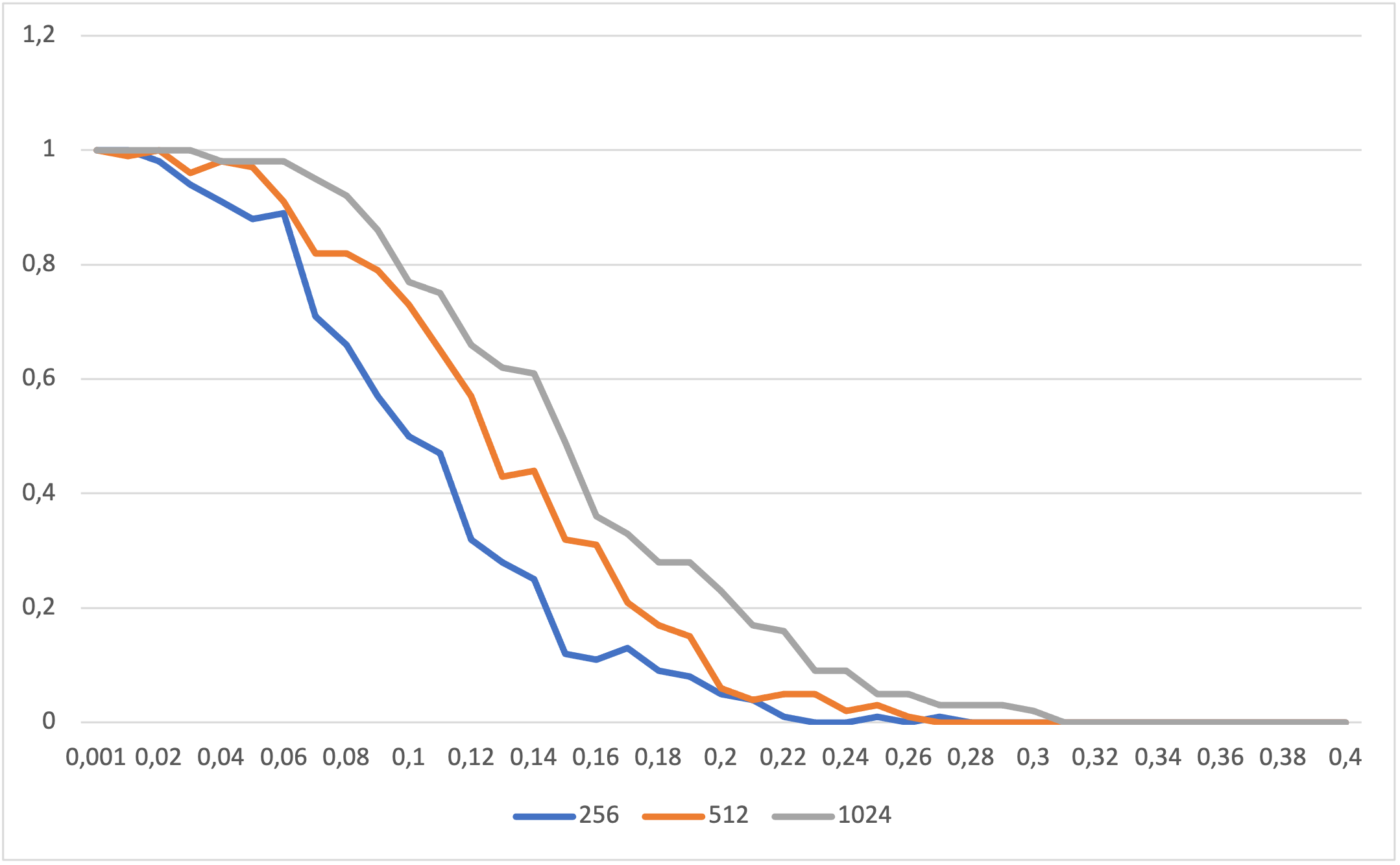}
         \caption{Full Enumeration for $\mu \in \{256,512,1024\}$}
         \label{fig:Picnic34_full}
     \end{subfigure}
     \end{adjustbox}
        \caption{Success rate of our key recovery algorithm with $W=192, w=8, \eta=3$, $\alpha=0.001$ and $\beta \in
        \{0.001, 0.01, 0.02, \ldots, 0.4\}$ for Picnic parameters \texttt{picnic-}\{\texttt{L3-FS}, \texttt{L3-UR}, \texttt{L3-full}\} and \texttt{picnic3-L3}. The $x$-axis represents $\beta$, while $y$-axis represents the success rate. }
        \label{fig:Picnic34}
\end{figure}

Based on the previous observation, we estimate the success rate of our key-recovery method by assuming the attacker can perform various enumerations from the set of candidates, $\mathcal{C}$, that can be constructed from $\texttt{L}$. In particular, we assume an attacker is able to enumerate (1) all candidates from $\mathcal{C}$, and (2) the $e$ best high-scoring candidates from $\mathcal{C}$, where $e \in \{2^{30}, 2^{40}, 2^{50}\}$ (this is basically what Algorithm~\ref{QKS} does for a given $e$).

\begin{figure}[ht!]
     \centering
     \begin{adjustbox}{minipage=\columnwidth}
     \begin{subfigure}{0.45\textwidth}
         \centering
         \includegraphics[width=\textwidth]{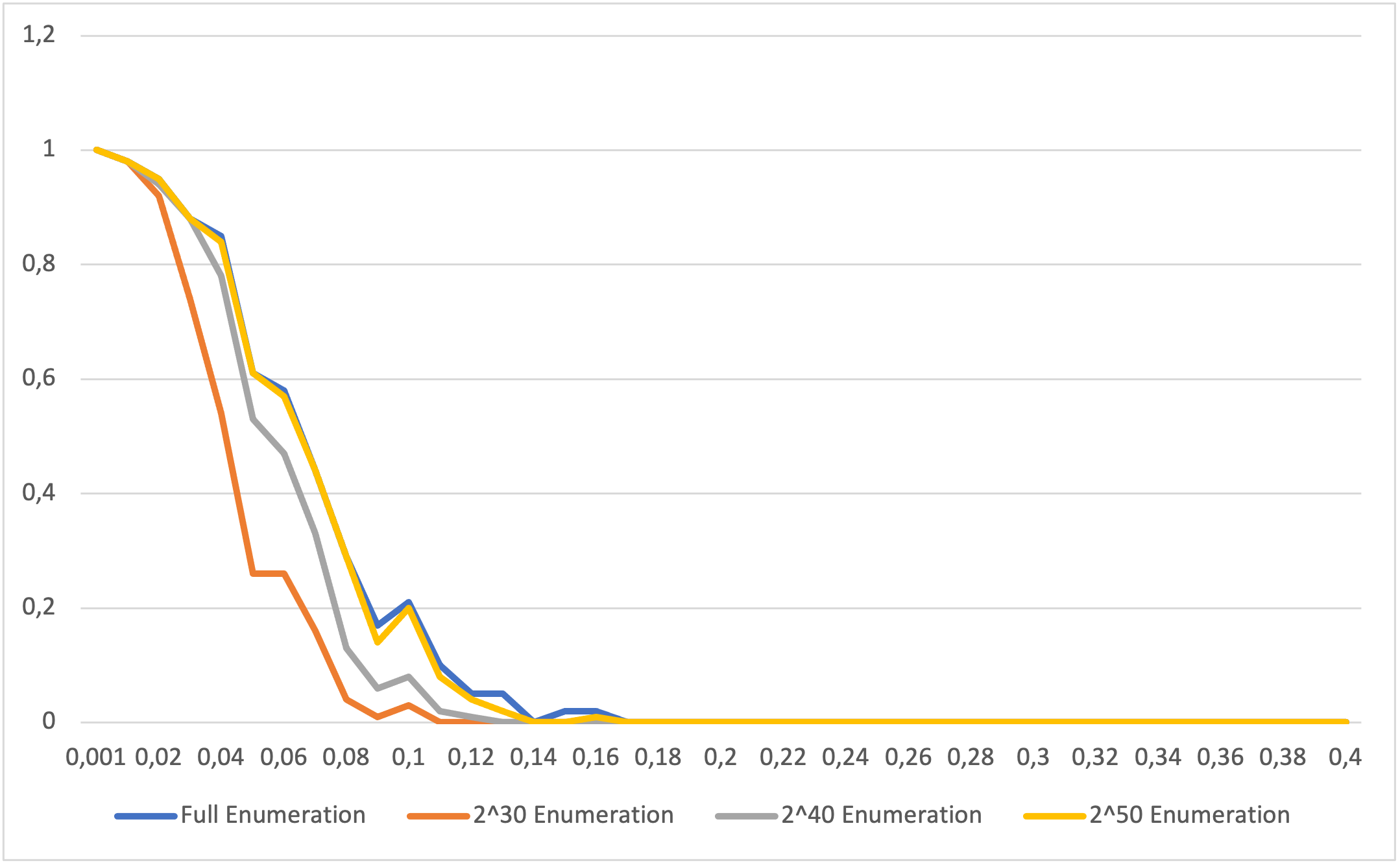}
         \caption{$\mu=256$}
         \label{fig:Picnic56_256}
     \end{subfigure}
     \hfill
     \begin{subfigure}{0.45\textwidth}
         \centering
         \includegraphics[width=\textwidth]{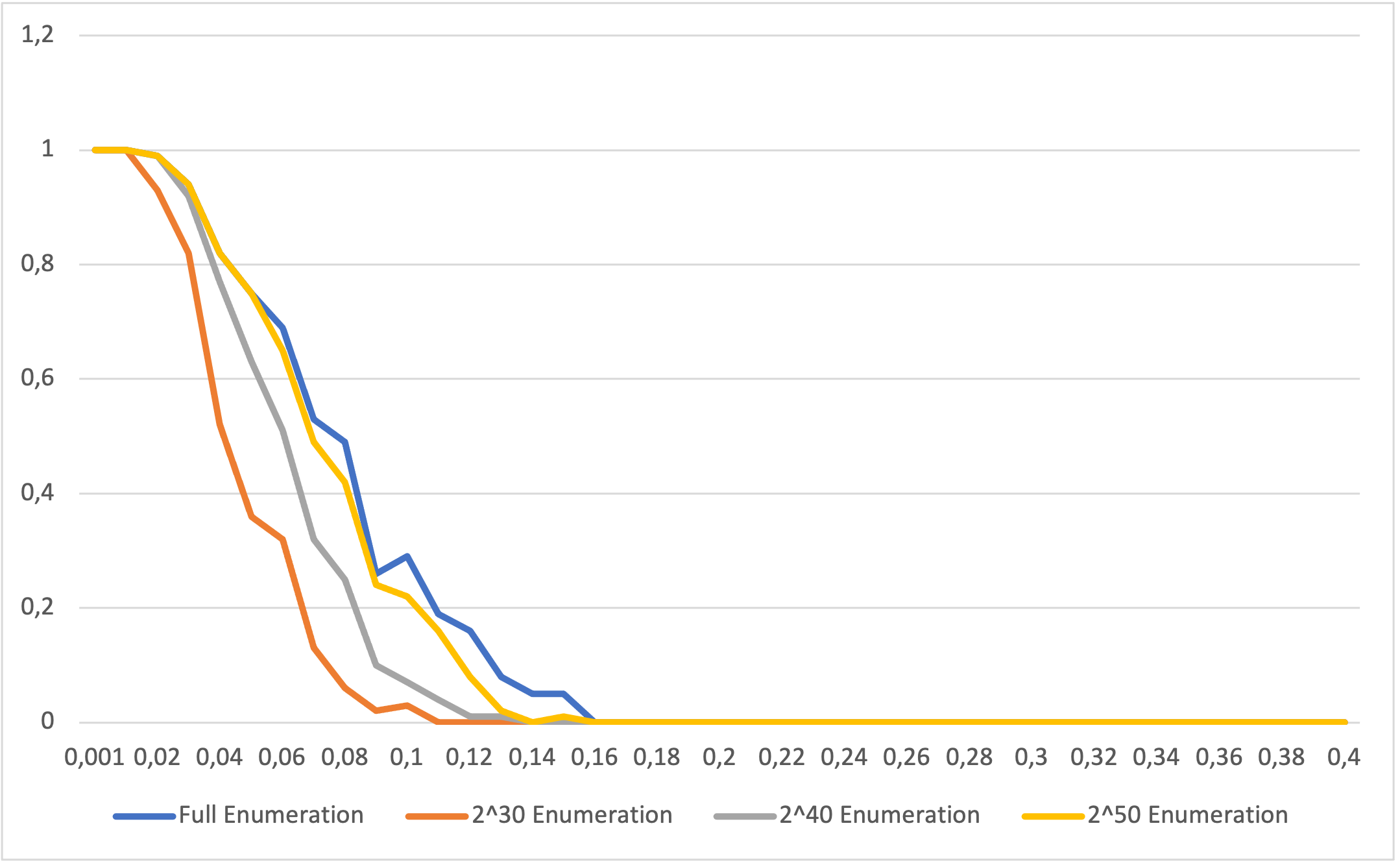}
         \caption{$\mu=512$}
         \label{fig:Picnic56_512}
     \end{subfigure}
     \begin{subfigure}{0.45\textwidth}
         \centering
         \includegraphics[width=\textwidth]{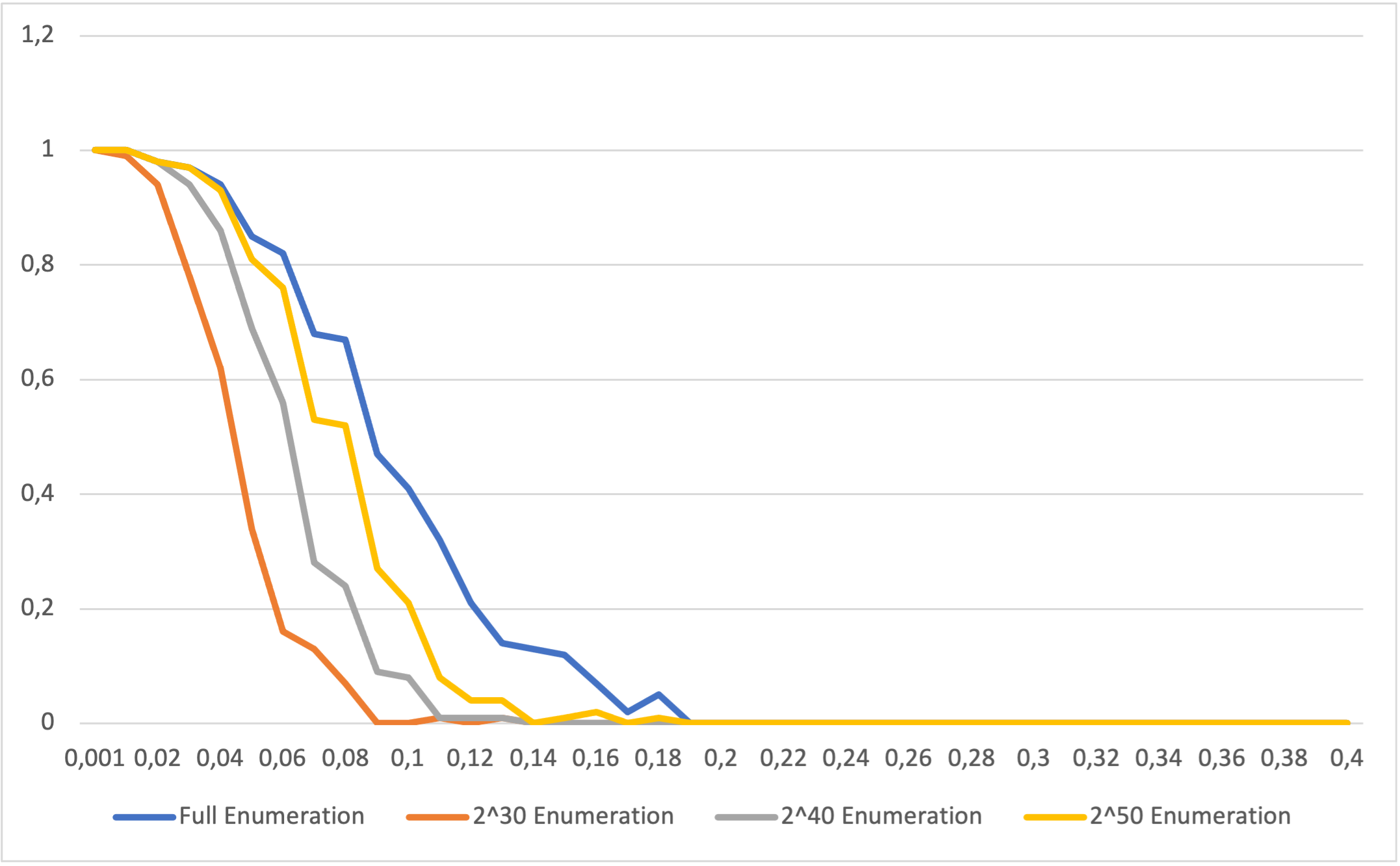}
         \caption{$\mu=1024$}
         \label{fig:Picnic56_1024}
     \end{subfigure}
     \hfill
     \begin{subfigure}{0.45\textwidth}
         \centering
         \includegraphics[width=\textwidth]{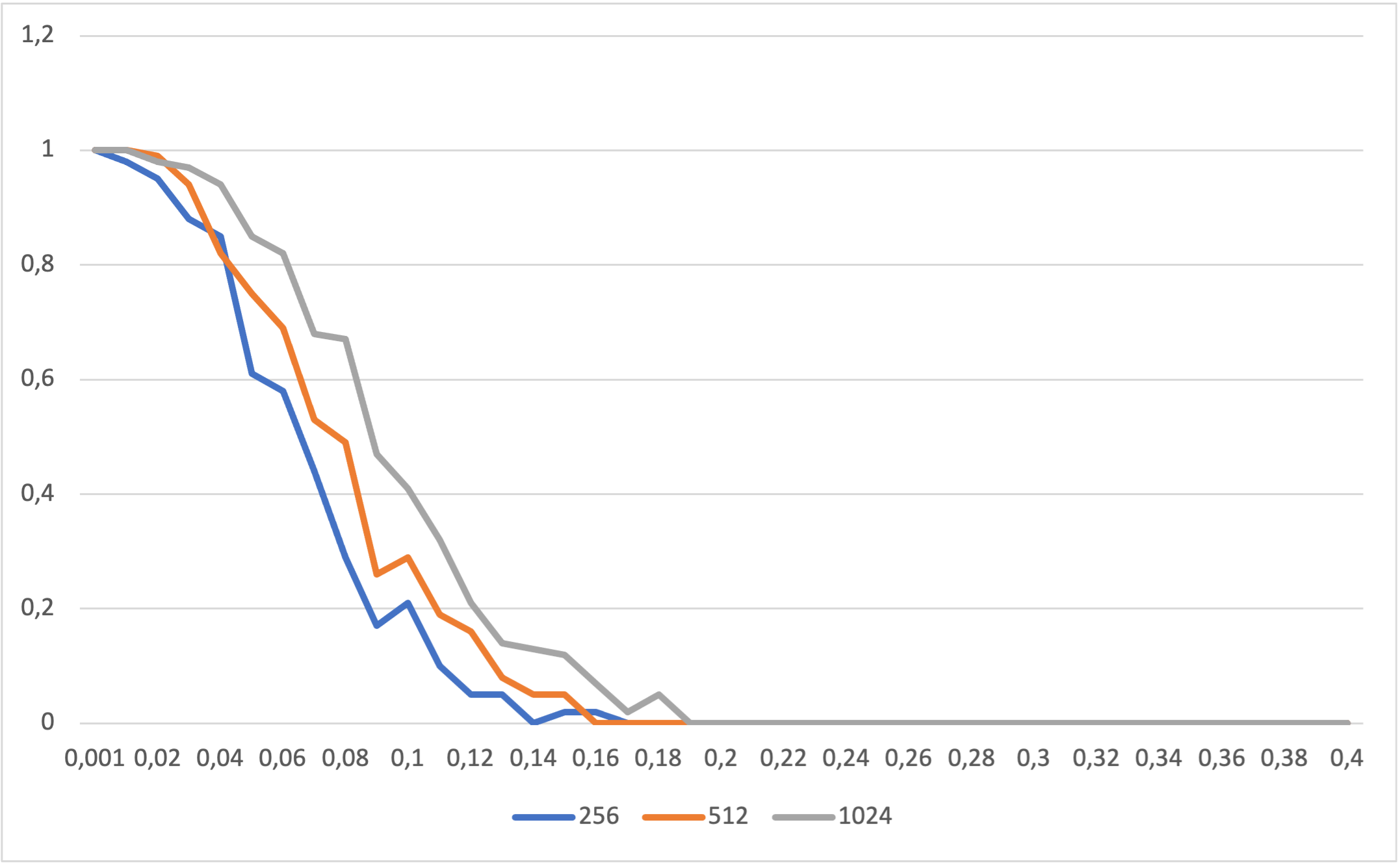}
         \caption{Full Enumeration for $\mu \in \{256,512,1024\}$}
         \label{fig:Picnic56_full}
     \end{subfigure}
     \end{adjustbox}
        \caption{Success rate of our key recovery algorithm with $W=256, w=8, \eta=4$, $\alpha=0.001$ and $\beta \in
        \{0.001, 0.01, 0.02, \ldots, 0.4\}$ for Picnic parameters \texttt{picnic-}\{\texttt{L5-FS}, \texttt{L5-UR}, \texttt{L5-full}\} and \texttt{picnic3-L5}. The $x$-axis represents $\beta$, while $y$-axis represents the success rate. }
        \label{fig:Picnic56}
\end{figure}

To calculate the success rate of our algorithm for a given $\alpha, \beta$ and a Picnic parameter set
$\texttt{P}$, we perform the following experiment that consists of $100$ trials. In each trial, we first create the key pair $ \texttt{sk},\texttt{pk}$ via calling the key generation algorithm from Picnic, as implemented in the Picnic reference implementation~\cite{pinicRI}. We then perturb \texttt{sk} according to $\alpha, \beta$ to get $\widetilde{sk}$. We then select appropriate values for $W, w,\eta,\mu$, and generate $\texttt{L}$ via calling Algorithm~\ref{generateCandidates} and check if the real key can be reconstructed from $\texttt{L}$, i.e., by verifying if the corresponding chunk candidates are in the lists of chunk candidates contained in $\texttt{L}$. If so, that signifies that a full enumeration can recover $\texttt{sk}$. Otherwise, \texttt{sk} cannot be recovered. Additionally, in case \texttt{sk} can be recovered by a full enumeration, we then calculate three intervals of the form $[B_{min}, B_e)$ for each $e$, as in Algorithm~\ref{QKS}, to check if the score of the real private key lies in each of them. Note that this check verifies if performing an enumeration of the $e$ best high-scoring candidates is enough to recover the real private key.

Figure~\ref{fig:Picnic12} shows the results for the Picnic parameters \texttt{picnic-}\{\texttt{L1-FS}, \texttt{L1-UR}, \texttt{L1-full}\} and \texttt{picnic3-L1}. In particular, it shows that our key recovery algorithm may find the real private key for $\alpha=0.001$ and $\beta$ in the set $\{0.001, 0.01, 0.02, \ldots, 0.4\}$ when run with the parameters $W=128, w=8, \eta=2$ and $\mu \in \{256,512,1024\}$. Note that the success rate  improves as the value of $e$ increases, which is expected. Similarly, Figure~\ref{fig:Picnic12_full} shows the success rate for the full enumeration improves as the the value of $\mu$ increases, which is also expected.
Additionally, note that although the bit length of the private key for the parameters sets \texttt{picnic-L1-full} and \texttt{picnic3-L1} is $129$ bits, the success rate of our algorithm for these two parameter sets is essentially the same as shown by Figure~\ref{fig:Picnic12}.

Figure~\ref{fig:Picnic34} shows the results for the Picnic parameters \texttt{picnic-}\{\texttt{L3-FS}, \texttt{L3-UR}, \texttt{L3-full}\} and \texttt{picnic3-L3}. In particular, it shows that our key recovery algorithm may find the real private key for $\alpha=0.001$ and $\beta$ in the set $\{0.001, 0.01, 0.02, \ldots, 0.3\}$ when run with the parameters $W=192, w=8, \eta=3$ and $\mu \in \{256,512,1024\}$. As mentioned before, the success rate  improves as the value of $e$ increases, which is expected. Similarly, Figure~\ref{fig:Picnic34_full} shows the success rate for the full enumeration improves as the the value of $\mu$ increases, which is also expected.

Figure~\ref{fig:Picnic56} shows the results for the Picnic parameters \texttt{picnic-}\{\texttt{L5-FS}, \texttt{L5-UR}, \texttt{L5}, \texttt{L5-full}\} and \texttt{picnic3-L5}. In particular, it shows that our key recovery algorithm may find the real private key for $\alpha=0.001$ and $\beta$ in the set $\{0.001, 0.01, 0.02, \ldots, 0.2\}$ when run with the parameters $W=256, w=8, \eta=4$ and $\mu \in \{256,512,1024\}$. As mentioned before, the success rate  improves as the value of $e$ increases, which is expected. Similarly, Figure~\ref{fig:Picnic56_full} shows the success rate for the full enumeration improves as the the value of $\mu$ increases, which is also expected. Additionally, note that although the bit length of the private key for the parameters sets \texttt{picnic-L5-full} and \texttt{picnic3-L5} is $255$ bits, the success rate of our algorithm for these two parameter sets is essentially the same as shown by Figure~\ref{fig:Picnic56}.

\section{Conclusions}
\label{sec:conclusions}

This paper presented a general procedure by which a cold boot attacker may recover a block cipher secret key after procuring a noisy version of the key via a cold boot attack. More specifically, the procedure exploits key enumeration algorithms and a well-known quantum algorithm, namely, Grover's Algorithm. Also, we showed how to implement the quantum component of our algorithm for several block ciphers such as AES, PRESENT and GIFT,and LowMC. This paper also evaluated Picnic, a post-quantum signature algorithm, in the cold boot attack setting, focusing on its reference implementation. We showed that our key-recovery method effectively reconstructs Picnic private keys for all Picnic parameters for $\alpha=0.001$ and values of $\beta$ in the set $\{0.001,0.01,0.02, \ldots, 0.4\}$ (the upper bound for $\beta$ depends on the used parameter set). Additionally, we provided the costs for running our key recovery algorithm by giving the number of quantum gates required to implement it and its running time. As future work, we believe that our key-recovery algorithm may be adapted to tackle key-recovery of other post-quantum algorithms' private keys in the cold boot attack setting.

\bibliographystyle{plain}
\bibliography{references}


\end{document}